\documentclass[twocolumn,showpacs,amsfonts,aps,prd,floatfix,nofootinbib,float,superscriptaddress]{revtex4}
\usepackage{amsmath}
\usepackage{bm}
\usepackage{graphicx}
\usepackage{color}
\usepackage{booktabs}
\usepackage{placeins}
\usepackage{enumerate}

\def\slashb#1{\not\!\!#1}
\def\delsla{\not\!\partial}

\begin{document}
\title{Heavy quark master equations in the Lindblad form at high temperatures}
\date{\today}

\author{Yukinao Akamatsu}
\affiliation{Kobayashi-Maskawa Institute for the Origin of Particles and the Universe (KMI), Nagoya University, Nagoya, Aichi 464-8602, Japan}

\begin{abstract}
We derive the quantum master equations for heavy quark systems in a high-temperature quark-gluon plasma in the Lindblad form.
The master equations are derived in the influence functional formalism for open quantum systems in perturbation theory.
These master equations have a wide range of applications, such as decoherence of a heavy quarkonium and Langevin dynamics of a heavy quark in the quark-gluon plasma.
We also show the equivalence between the quarkonium master equations in the recoilless limit and the Schr\"odinger equations with stochastic potential.
\end{abstract}


\maketitle

\section{Introduction}
\label{sec:introduction}
The fate of heavy quarkonium bound states in finite-temperature QCD matter has long been considered as a sensitive probe of deconfined nature of such a matter.
In the deconfined phase at finite temperature, the linear potential that confines a heavy quark-antiquark pair in the vacuum is screened by colored excitations (light quarks and gluons) in the medium and in such a short-ranged potential the quarkonium bound state levels will eventually disappear at high temperature.
In relativistic heavy-ion collisions, suppression of the quarkonium yield, in particular $\Upsilon$ and $J/\Psi$ states, is expected to serve as a signal for the formation of a quark-gluon plasma (QGP) \cite{Matsui:1986dk}, the deconfined matter that once existed just after the big bang in the early Universe.
Indeed, experimental data from the CMS collaboration at the LHC show sequential suppression of $\Upsilon$ states [the ground (1S) and excited states (2S, 3S)] \cite{Chatrchyan:2011pe}, suggesting sequential melting of the $\Upsilon$ bound states in the QGP.
To investigate the real-time dynamics of quarkonium quantum states and its suppression in the QGP, the appropriate theoretical framework is that of open quantum systems \cite{BrePetText}.

There have been various studies on the quarkonium properties at finite temperature.
Thermodynamic quantities, such as free energy change caused by putting an infinitely heavy quark and antiquark pair, are calculated in lattice QCD simulations.
The results clearly show that color charges are screened above the deconfinement transition temperature $T_{\rm c}$ \cite{Petreczky:2012rq}.
Spectral structures of $\Upsilon$ and $J/\Psi$ are investigated by lattice QCD simulations and suggest that the ground states are fairly stable even at higher temperature up to $T<2T_{\rm c}$. The stability of quarkonium ground states indicates a strongly coupled nature of the quark-gluon plasma \cite{Aarts:2010ek,Asakawa:2003re}.
It is not yet clear how these independent observations by numerical simulations can be understood in a unified point of view.

Recently, a real-time static potential, defined in terms of the real-time propagator of the quarkonium operator at finite temperature, has been calculated in perturbation theory \cite{Laine:2006ns}, nonperturbative lattice QCD simulations \cite{Rothkopf:2011db}, and the potential nonrelativistic QCD approach \cite{Brambilla:2008cx}.
The same quantity has also been calculated for strong coupling plasmas using the conjectured gauge/gravity duality \cite{Albacete:2008dz}.
The real-time static potential is one of the crucial quantities to understand the quarkonium dynamics in the QGP.
The potential is found to be complex valued with a negative imaginary part.
Using this complex-valued potential, one can calculate the spectral functions for quarkonia in the QGP \cite{Laine:2007gj}.
Although the potential has an imaginary part, particle number conservation of the nonrelativistic heavy quarks with infinite mass is not violated.
Quantum decoherence of quarkonium wave functions due to stochastic processes (the stochastic potential) can give a physically natural explanation to the imaginary part \cite{Akamatsu:2011se}.
Clearly, the complex potential and its stochastic potential interpretation indicates that the quantum mechanical properties of quarkonium in the QGP must be studied from the viewpoint of the open quantum systems \cite{BrePetText}, which we will summarize briefly.

\subsection{Basics of open quantum systems}
In general, dynamics of open quantum systems is characterized by the reduced density matrix $\hat\rho_{\rm S}(t)$ defined as
$\hat\rho_{\rm S}(t)\equiv {\rm Tr_{\rm E}}\hat\rho_{\rm tot}(t)$.
Here $\hat\rho_{\rm tot}(t)$ is the total density matrix for both system and environment degrees of freedom and $\rm Tr_{\rm E}$ denotes the trace over the environment degrees of freedom.
In the case of heavy quark systems in the QGP, the system consists of heavy quarks and the environment is composed of light quarks and gluons.
Time evolution of $\hat\rho_{\rm S}(t)$ in the Markov limit is given by the quantum master equation:
\begin{eqnarray}
\frac{d}{dt}\hat\rho_{\rm S}(t)={\mathcal L}\hat\rho_{\rm S}(t).
\end{eqnarray}
Note that the generator of the time evolution $\mathcal L$ is a superoperator which acts linearly on the operator $\hat\rho_{\rm S}(t)$.

There are two major regimes in the open quantum systems \cite{BrePetText}:
Quantum optical limit and quantum Brownian motion.
The former applies to systems whose intrinsic time scale $\tau_{\rm S}$ is much shorter than their relaxation time $\tau_{\rm R}$.
In the quantum optical limit, one can distinguish quantum states with typical energy level differences ($\tau_{\rm S}^{-1}$) in the time scale of interest ($\tau_{\rm R}$), and the so-called rotating wave approximation is applicable.
The latter applies to systems where $\tau_{\rm S}$ is much longer than the correlation time of the environment $\tau_{\rm E}$.
In the quantum Brownian motion, $\tau_{\rm S}$ is estimated by the orbital period of the Brownian particle.
When $\tau_{\rm E}\ll\tau_{\rm S}$, one can neglect the acceleration of the Brownian particle during a short time period $\tau_{\rm E}$.
In both of these regimes, $\tau_{\rm E}\ll\tau_{\rm R}$ must be satisfied so that the system is insensitive to the initial condition of the environment.
The conditions on the time scales are summarized in Table \ref{tab:oqs_regimes}.
Roughly, these two regimes correspond to different choices of the unperturbed Hamiltonian and representation basis of quantum states (such as eigenstate basis or position-space basis).

\begin{table}[top]
\begin{tabular}{c|c}
\hline
 Quantum optical limit & Quantum Brownian motion \\
\hline
$\tau_{\rm S}\ll\tau_{\rm R}$ & $\tau_{\rm E}\ll\tau_{\rm S}$ \\
$\tau_{\rm E}\ll\tau_{\rm R}$ & $\tau_{\rm E}\ll\tau_{\rm R}$ \\
\hline
\end{tabular}
\caption{Two major regimes of open quantum systems \cite{BrePetText}.}
\label{tab:oqs_regimes}
\end{table}

Several studies have applied the open quantum system descriptions to heavy quark systems \cite{Young:2010jq,Borghini:2011yq,Akamatsu:2011se,Akamatsu:2012vt}.
In terms of the two regimes of the open quantum systems, Ref.~\cite{Borghini:2011yq} considers the quantum optical limit while Refs.~\cite{Young:2010jq,Akamatsu:2011se,Akamatsu:2012vt} correspond to the quantum Brownian motion.
In Ref.~\cite{Borghini:2011yq}, the quantum optical description for a quarkonium is derived, but the treatment of unbound color octet states is rather obscure.
Reference \cite{Young:2010jq} demonstrates how the heavy quark quantities, such as mass, potential, and drag force, affect the imaginary-time current correlator, assuming the Caldeira-Leggett model \cite{Caldeira:1982iu} for quantum Brownian motion of heavy quarks.
In Ref.~\cite{Akamatsu:2012vt}, quantum master equations are first derived at leading order in perturbation for nonrelativistic heavy quark systems.
In this derivation, the Feynman-Vernon's influence functional formalism \cite{Feynman:1963fq} is applied to the finite-temperature QCD.

\subsection{Summary of main results}
One of the purposes of this paper is to extend the result of Ref.~\cite{Akamatsu:2012vt} and derive heavy quark master equations in the Lindblad form \cite{Lindblad:1975ef}.
In particular, we derive explicit forms of the Lindblad-form master equations for a single heavy quark system and for a heavy quark-antiquark system in the regime of quantum Brownian motion.
The Lindblad form is the form of the superoperator $\mathcal L$ that any Markovian master equation that preserves complete positivity of the reduced density matrix must conform to.
The Lindblad form is generally expressed with a Hermitian Hamiltonian $\hat H$, Lindblad operators $\hat L_i$, and positive coefficients $\gamma_i>0$ $(i=1,2,\cdots,N)$:
\begin{eqnarray}
\label{eq:Lindblad1}
\frac{d}{dt}\hat\rho_{\rm S}(t)&=&-i[\hat H,\hat \rho_{\rm S}] \\
&+&\sum_{i=1}^{N}\gamma_i\left(
\hat L_i \hat \rho_{\rm S} \hat L^{\dagger}_i
-\frac{1}{2} \hat L_i^{\dagger} \hat L_i \hat \rho_{\rm S}
-\frac{1}{2}\hat \rho_{\rm S} \hat L_i^{\dagger} \hat L_i
\right).\nonumber
\end{eqnarray}
Here, $N$ is not necessarily connected to the dimension of the Hilbert space.
We do not know $\hat H$, $\hat L_i$, $\gamma_i$, and $N$ {\it a priori}.
By deriving the master equations in the Lindblad form, we may be able to utilize several techniques, such as the quantum state diffusion method \cite{Gisin:1992} and quantum jump method \cite{Plenio:1997ep}, to numerically simulate the master equation in terms of wave function.
In general, numerical calculation with a wave function has substantial advantage over that of the master equation because the dimension of a wave function is the square root of that of a density matrix.

By applying the influence functional formalism \cite{Feynman:1963fq} to QCD at finite temperature, the master equations in the Lindblad form are derived from an influence functional with proper order of time coarse graining
\begin{eqnarray}
S_{\rm IF}=S_{\rm pot}+S_{\rm fluct}+S_{\rm diss}+S_{\rm L}.
\end{eqnarray}
A definition of the influence functional $S_{\rm IF}$ and explicit forms of each term will be given in Sec.~\ref{sec:influence_functional} and Eqs.~\eqref{eq:IF_pot}-\eqref{eq:IF_fluct} and \eqref{eq:IF_diss2}-\eqref{eq:IF_acc2}.
For each of the master equation, we will explicitly identify the operators and parameters in the Lindblad form $\hat H$, $\hat L_i$, $\gamma_i$.
We find that inclusion of $S_{\rm L}$ is essential in obtaining the Lindblad-form master equations.

The other purpose is to present a theoretical basis to the concept of stochastic potential, which was first introduced in Ref.~\cite{Akamatsu:2011se} and has been recently simulated in Ref.~\cite{Rothkopf:2013kya}.
This is partly because we find several confusing applications of the complex potential to the problem of quarkonium survival probability.
\footnote{
(i) There is a conceptual problem if one calculates expectation values by using a wave function that is evolved by the Schr\"odinger equation with the complex potential and its conjugate \cite{Margotta:2011ta,CasalderreySolana:2012av}.
(ii) Using the complex potential, one can calculate the width \cite{Brambilla:2010vq}.
But the width only gives a rate of transition from one state to any of the other states in one scattering, which is sometimes insufficient to describe the dynamics.
}
By definition, the master equation corresponding to the stochastic potential is of the Lindblad form because we derive it from the ensemble of wave functions with positive probability.
Therefore, the stochastic potential can be regarded as a method to calculate certain types of the Lindblad-form master equations in terms of wave function.
The stochastic potential has two sources of quantum decoherence:
One is decoherence among the wave functions in the ensemble at the same point $X=(\vec x_Q,\vec x_{Q_c})$, and the other is decoherence in each wave function at different points $X$ and $Y$, where $\vec x_Q$ and $\vec x_{Q_c}$ denote positions of heavy quark and antiquark \cite{Akamatsu:2011se}. 
With the information of the complex potential, we only know the former source for decoherence.
Thus, if we only know the complex potential, as is the case at present for nonperturbative lattice calculation of the complex potential \cite{Rothkopf:2011db}, we can just guess the latter, for example, by referring to the perturbative results.
In the perturbative analysis, we see that the imaginary part of the complex potential has enough information to know the decoherence at different points $X$ and $Y$.

In terms of the influence functional $S_{\rm IF}=S_{\rm pot}+S_{\rm fluct}+S_{\rm diss}+S_{\rm L}$ above, the stochastic potential derives from $S_{\rm pot}$ and $S_{\rm fluct}$.
The resultant stochastic Schr\"odinger equation is given in \eqref{eq:QQc_stochastic_rel}, which we quote here
\begin{eqnarray}
&&i\frac{\partial}{\partial t}\psi^r(t,\vec r)
=H^r_{\xi}(t)\psi^r(t,\vec r),\\
\label{eq:QQc_stochastic_rel_Hamiltonian}
&&H^r_{\xi}(t)= -\frac{\vec{\nabla}_r^2}{M}+iC_{\rm F}D(\vec 0)
+\left(-V(\vec r)-iD(\vec r)\right)(t^a\otimes t^{a*})\nonumber\\
&&\ \ \ \ \ \ \ \ + \ \theta^a(t,\vec r/2)(t^a\otimes 1) - \theta^a(t,-\vec r/2)(1\otimes t^{a*}).
\end{eqnarray}
Here $\psi^r(t,\vec r)$ is a quarkonium wave function in $N_{\rm c}\otimes N_{\rm c}^*$ representation of the color ${\rm SU}(N_{\rm c})$ group and $\theta^a$ is a white noise with color (see the main text for more details).
The random color rotation by the stochastic potential is a unique feature in the quark-gluon plasma.
We also discuss quantum decoherence for a bound state of size $l_{\rm coh}$ and estimate typical time scales for the decoherence as in Eq.~\eqref{eq:decoh_timescale}:
\begin{eqnarray}
t_{\rm D}(l_{\rm coh},T)\sim \frac{1}{g^2T}\left(a+\frac{b}{g^2\ln (1/g)T^2l_{\rm coh}^2}\right),
\end{eqnarray}
with $a$ and $b$ of order $\mathcal O(g^0)$.
Clearly, it takes a longer time for smaller bound states to get decoherent and excited.

This paper is organized as follows.
In Sec.~\ref{sec:influence_functional}, we begin with a review of the method developed in Ref.~\cite{Akamatsu:2012vt} and update it by including a new term necessary to obtain the master equations in the Lindblad form.
In Sec.~\ref{sec:master_equations}, we derive several master equations for a single heavy quark and for a quarkonium in the QGP.
We show that the master equations can be simplified for localized wave packets.
We also show that if the coherence length of a wave function is long enough, decoherence phenomena can be described by master equations in the recoilless limit.
Each of them is shown to be in the Lindblad form.
In Sec.~\ref{sec:decoherence}, we give the stochastic potential with color degrees of freedom, an extension of Ref.~\cite{Akamatsu:2011se}.
We then study the decoherence of a quarkonium wave function by comparing two scales, correlation length of thermal fluctuation and coherence length of the wave function.
We also discuss how quantum wave function description can be evolved into a classical regime through decoherence.
Section \ref{sec:summary} is devoted to a summary.
Throughout this paper, we adopt the natural units, $\hbar=c=k_{\rm B}=1$, and operators in Hilbert and Fock spaces are denoted by bold fonts.

\section{Influence functional of heavy quarks}
\label{sec:influence_functional}
In this section, we review and also update the formalism developed in Ref.~\cite{Akamatsu:2012vt}.
The formalism relies on three approximations for actual computations:
(i) A nonrelativistic limit of heavy quarks $v\ll 1$,
(ii) perturbative expansion in terms of coupling constant $g\ll 1$, and
(iii) coarse graining in time.
Since the heavy quarks are nonrelativistic, we only consider the color density interaction in the couplings between heavy quarks and gluons, which is the leading contribution in the $1/c$ expansion \cite{ManoharText}.
The region of validity of these approximations is summarized in Table \ref{tab:validity}.
Through these approximations, we can obtain the influence functional and renormalized effective Hamiltonian, from which master equations for an arbitrary number of heavy quarks can be derived.

\begin{table}[top]
\begin{tabular}{l |c|c}
\hline
 & Heavy quark & Heavy quarkonium \\
\hline
(i) Nonrelativistic limit & $\sqrt{T/M}\ll1$ & $\alpha,\sqrt{T/M}\ll 1$ \\
(ii) Perturbation & $g\ll1$ & $g\ll 1$ \\
(iii) Coarse graining & $-$ & $1/gT \ll 1/M\alpha^2$ \\
\hline
\end{tabular}
\caption{
Region of validity of the approximations.
The validity for the quarkonium case is evaluated assuming the Coulomb bound states.
Also we only consider the leading contribution of $1/c$ expansion in the heavy quark interactions.
}
\label{tab:validity}
\end{table}

\subsection{Heavy quark velocity and acceleration}
Here we consider heavy quarks and quarkonium bound states close to their kinetic equilibrium.
As we will see, heavy quark velocity in such a condition is small in the rest frame of the thermal medium.
However, in realistic situations in the heavy-ion collisions, they are not always close to kinetic equilibrium and thus the nonrelativistic approximation is sometimes not appropriate for phenomenological studies.
For quarkonium bound states, in addition to the velocity, we also need to estimate the acceleration by potential force in order to make coarse graining in time.

First of all, the heavy quark velocity in unbound states is estimated as $v\sim\sqrt{T/M}$ close to kinetic equilibrium.
Here the heavy quark mass is $M_{\rm b}\approx 4.8$ GeV and $M_{\rm c}\approx 1.5$ GeV for bottom and charm quarks and the typical temperature is $T\approx(1-3) T_{\rm crit}\sim 200-500$ MeV in the heavy-ion collision experiments.
Therefore, close to kinetic equilibrium, the heavy quark velocity is small for both bottom and charm quarks in unbound states.

In the case of quarkonium bound states, we also expect that the quarkonium velocity is small close to equilibrium.
In addition, we need to take into account the relative velocity and acceleration of a heavy quark-antiquark pair.
Let us now consider bound states in the Coulomb potential $V(r)=-\alpha/r$ since the fastest relative velocity can be estimated by the most deeply bound states.
The Coulomb part of the phenomenological Cornel potential is typically chosen as $\alpha\sim 1/4$ \cite{Bali:2000gf}.
The momentum of Coulomb bound states is estimated as $p\sim M\alpha$ and thus $v\sim\alpha\sim 1/4$. 
Therefore, in the temperature range of phenomenological interest, we can assume that the relative velocity of the heavy quark-antiquark pair is small.

As for the acceleration by the potential force, it can be estimated by $\dot v\sim \alpha/Mr^2\sim M\alpha^3$.
When we perform coarse graining in time later, we need to assume that the effect of acceleration is small during a scattering event.
Since typical correlation time of the medium is $\sim 1/gT$ or shorter ($\sim 1/T$), the condition is obtained as $M\alpha^2 \ll gT$.
This condition corresponds to $\tau_{\rm E}\ll\tau_{\rm S}$ for the quantum Brownian motion in Table \ref{tab:oqs_regimes}.
Using phenomenological values $\alpha\sim 1/4$ and $g\sim 2$, this condition is satisfied for charmonium but is not very obvious for bottomonium at lower temperature.
Nevertheless, we neglect the effect of acceleration in the bottomonium bound states in the coarse-graining procedure because $M\alpha^2 \gg gT$ is also not at all obvious for these states.
\footnote{
A relation $\tau_{\rm E}\approx \tau_{\rm S}$ indicates that the system should be treated in the quantum optical limit, where the system is described with a few relevant bound states.
If one attempts to obtain such a description, one would need to evaluate transition amplitudes between those singlet bound states.
}

\subsection{Influence functional}
The influence functional can be defined using the closed-time path formalism of nonequilibrium field theory \cite{Akamatsu:2012vt}.
In the closed-time path formalism \cite{Schwinger:1960qe}, fields $\phi =(A,q,\psi)$ on the forward (backward) time axis are denoted by $\phi_1(\phi_2)$, where $A$ is the gauge field, $q$ is the light quark field, and $\psi$ is the heavy quark field.
Since the time-integration contour is originally a closed path, the fields $\phi_1$ and $\phi_2$ satisfy proper boundary conditions at $t\to\infty$.
The sources $\eta_{1,2}$ for the fields $\phi_{1,2}$ are also introduced.
The partition function of the total system is defined as
\begin{eqnarray}
Z[\eta_1,\eta_2]
&=&\int \mathcal {D}[\phi]_{1,2}
\langle \phi_1 |{\bm\rho}_{\rm tot}|\phi_2 \rangle_{t_0} \\
&& \ \times 
\exp\Bigl[i\int_{t_0} d^4x\left\{
\mathcal{L}_{\rm tot}(\phi_1)-\phi_1\eta_1
\right\}\Bigl]\nonumber\\
&& \ \times 
\exp\Bigl[-i\int_{t_0} d^4x\left\{
\mathcal{L}_{\rm tot}(\phi_2)-\phi_2\eta_2
\right\}\Bigl],\nonumber
\end{eqnarray}
where $\mathcal{L}_{\rm tot}(\phi)$ denotes the Lagrangian density for the total system of gluons, light quarks, and heavy quarks.
Here, $|\phi\rangle$ is the coherent state introduced to obtain path-integral formulation.
The contributions from the gauge fixing term and ghost are implicit here.

Let us assume that the initial density matrix for the total system is factorized as ${\bm\rho}_{\rm tot}={\bm\rho}^{\rm eq}_{\rm E}\otimes{\bm\rho}_{\rm S}$ with ${\bm \rho}_{\rm E}^{\rm eq}$ being the equilibrium density matrix at temperature $T$ for interacting gluons and light quarks.
Switching off the sources, we arrive at 
\begin{eqnarray}
Z[0,0]&=&\int \mathcal {D}[\psi]_{1,2}
\langle \psi_1^{\dagger} |{\bm\rho}_{\rm S}|\psi_2 \rangle_{t_0}\\
&&\times \exp\left[{iS_{\rm kin}[\psi_1]-iS_{\rm kin}[\psi_2]+iS_{\rm IF}[j_1,j_2]}\right], \nonumber
\end{eqnarray}
where the influence functional $S_{\rm IF}$ is defined as a functional of the heavy quark color current $j^{a\mu}=\bar\psi t^a\gamma^{\mu}\psi$:
\begin{eqnarray}
e^{iS_{\rm IF}[j_1,j_2]}&=&
\int \mathcal {D}[A, q]_{1,2}
\langle A_1,q_1 |{\bm\rho}^{\rm eq}_{\rm E}|A_2,q_2 \rangle_{t_0} \\
&& \times\exp
\Bigl[i\int_{t_0} d^4x\left\{
\mathcal{L}_{g+q}(A_1, q_1)-gj^{a\mu}_1A^a_{1\mu}
\right\}\Bigl]\nonumber \\
&& \times\exp
\Bigl[-i\int_{t_0} d^4x\left\{
\mathcal{L}_{g+q}(A_2, q_2)-gj^{a\mu}_2A^a_{2\mu}
\right\}\Bigr].\nonumber
\end{eqnarray}
Here, $S_{\rm kin}(\psi)$ is the kinetic term for heavy quarks and $\mathcal{L}_{g+q}(A,q)$ is the Lagrangian density for gluons and light quarks.
As we see later, the influence functional provides time evolution of the reduced density matrix $\langle\psi_1^{\dagger}|{\bm\rho}_{\rm S}(t)|\psi_2\rangle$.

\subsubsection{Nonrelativistic limit}
Since heavy quark velocity is small, we take a nonrelativistic limit for the heavy quark Lagrangian.
Here, we take the leading order of the $1/c$ expansion \cite{ManoharText}, where $c$ is the velocity of light.
By recovering $c$ to distinguish the time and length scales, the heavy quark Lagrangian becomes
\begin{eqnarray}
\mathcal{L}_{\psi}
&=&c\bar\psi\left(i\delsla - \frac{g}{c}\slashb{A}^at^a - Mc\right)\psi\nonumber\\
&=&Q^{\dagger} \left[i(\partial_t + igA^a_0 t^a )+ \frac{\nabla^2}{2M} + \cdots \right]Q\nonumber\\
&& + \ Q_c \left[i(\partial_t + igA^a_0 t^a) - \frac{\nabla^2}{2M} + \cdots \right]Q_c^{\dagger},
\end{eqnarray}
where the expansion continues with $\mathcal O(1/c)$ corrections.
Here we take $\partial_0=\partial_t/c$ and $Q$ and $Q_c$ are heavy quark and antiquark fields, respectively, in the nonrelativistic limit.
Keeping only the leading-order terms in the $1/c$ expansion, the heavy quark Lagrangian contains only the nonrelativistic kinetic term and the interaction term with gluon scalar potential ($A^a_0$).
This expansion explicitly breaks the full gauge invariance but still there remains an invariance with respect to the temporal gauge transformation.
\footnote{
One can easily check that the resultant master equations give gauge invariant expectation values for color singlet observables.
Here, the gauge transformation is, of course, limited to the temporal direction $\psi(t,\vec x)\to U(t)\psi(t,\vec x)$ with $U(t)\in {\rm SU}(N_{\rm c})$.
}
Note that the $1/c$ expansion for the heavy quark Lagrangian does not necessarily lead to the same expansion in the final result.
This is because the environment is still relativistic and inevitably involves factors of $c$ if recovered.
If we formally distinguish the velocity of light in the heavy quark Lagrangian by using $c_{\rm Q}$, the nonrelativistic limit here corresponds to taking $1/c_{\rm Q}\to 0$.
Despite these shortcomings, we take this approach because of a clear physical picture: Heavy quarks interact with the environment through the color electric interaction.
\footnote{
In the $1/M$ expansion, the full gauge invariance is preserved in the heavy quark Lagrangian.
In this expansion, heavy quarks interact with the environment through the color magnetic interaction as well as the color electric interaction.
}
The influence functional now becomes a functional of heavy quark color density $\rho^a=\bar\psi t^a\gamma^0\psi$:
\begin{eqnarray}
e^{iS_{\rm IF}[\rho_1,\rho_2]}&\simeq&
\int \mathcal {D}[A, q]_{1,2}
\langle A_1,q_1 |{\bm\rho}^{\rm eq}_{\rm E}|A_2,q_2 \rangle_{t_0} \\
&& \times\exp
\Bigl[i\int_{t_0} d^4x\left\{
\mathcal{L}_{g+q}(A_1, q_1)-g\rho^{a}_1A^a_{1,\mu=0}
\right\}\Bigl]\nonumber \\
&& \times\exp
\Bigl[-i\int_{t_0} d^4x\left\{
\mathcal{L}_{g+q}(A_2, q_2)-g\rho^{a}_2A^a_{2,\mu=0}
\right\}\Bigr].\nonumber
\end{eqnarray}

\subsubsection{Perturbative expansion}
In the perturbative expansion, assuming the medium temperature is very high but much lower than the heavy quark mass, the leading-order terms in $S_{\rm IF}$ are given by
\begin{eqnarray}
\label{eq:IF_perturbation}
iS_{\rm IF}\left[\rho_1,\rho_2\right]
&=&-\frac{g^2}{2}\int_{t_0} d^4x d^4y
\left(
\rho^{a}_1,\ \rho^{a}_2
\right)_{(x)}\\
&\times&\left[
\begin{array}{cc}
G^{\rm F}_{ab,00}
 & -G^<_{ab,00} \\
-G^>_{ab,00}
 & G^{\rm \tilde F}_{ab,00}
\end{array}
\right]_{(x-y)}
\left(
\begin{array}{c}
\rho^{b}_1\\
\rho^{b}_2
\end{array}
\right)_{(y)}\nonumber \\
&+&\mathcal O(g^3).\nonumber
\end{eqnarray}
The two-point functions of gluons are defined as
\begin{eqnarray}
G^{\rm F}_{ab,00}(x-y)
&\equiv& \langle{\rm T}{\bm A}^a_{0}(x){\bm A}^b_{0}(y)\rangle,  \\
G^{\rm \tilde F}_{ab,00}(x-y)
&\equiv& \langle{\rm \tilde T}{\bm A}^a_{0}(x){\bm A}^b_{0}(y)\rangle, \\
G^>_{ab,00}(x-y)
&\equiv& \langle {\bm A}^a_{0}(x){\bm A}^b_{0}(y)\rangle, \\
G^<_{ab,00}(x-y)
&\equiv& \langle {\bm A}^b_{0}(y){\bm A}^a_{0}(x)\rangle,
\end{eqnarray}
where $\langle\mathcal O\rangle$ denotes the thermal average in the gluon and light quark system.
For completeness, let us also define the following retarded and advanced propagators, symmetrized correlation function, and spectral function:
\begin{eqnarray}
G^{\rm R}_{ab,00}(x-y)
&\equiv& i\theta(x^0-y^0)\langle\left[{\bm A}^a_{0}(x),{\bm A}^b_{0}(y)\right]\rangle,  \\
G^{\rm A}_{ab,00}(x-y)
&\equiv& -i\theta(y^0-x^0)\langle\left[{\bm A}^a_{0}(x),{\bm A}^b_{0}(y)\right]\rangle,\\
G^{\rm S}_{ab,00}(x-y)
&\equiv& \langle\left\{{\bm A}^a_{0}(x),{\bm A}^b_{0}(y)\right\}\rangle,\\
\sigma_{ab,00}(\omega,\vec x-\vec y)&\equiv& \int dte^{-i\omega(x^0-y^0)}
\langle\left[{\bm A}^a_{0}(x),{\bm A}^b_{0}(y)\right]\rangle.\nonumber \\
\end{eqnarray}

For a later purpose of coarse graining in time, let us change the time variables from $(x^0,y^0)$ to $(t,s)$ with
\begin{eqnarray}
t={\rm max}(x^0,y^0), \ s=|x^0-y^0|.
\end{eqnarray}
The new time variable $t$ is taken to be always the later one of $x^0$ and $y^0$.
This is essential in obtaining correct time-evolution equations.
In terms of the new time variables, the interaction terms can be schematically written as
\begin{eqnarray}
&& \int_{t_0} d^4x d^4y\rho(x)G(x-y)\rho(y)\\
&& = \int_{t_0}^{\infty} dt\int_0^{t-t_0} ds\int d^3xd^3y\nonumber \\
&& \ \ \ \ \ \ \times\left(
\begin{array}{l}
 \rho(t,\vec x)G(s,\vec x-\vec y)\rho(t-s,\vec y) \\
+\rho(t-s,\vec x)G(-s,\vec x-\vec y)\rho(t,\vec y)
\end{array}
\right)\nonumber \\
&& \simeq \int_{t_0}^{\infty} dt\int_0^{\infty} ds\int d^3xd^3y\nonumber \\
&& \ \ \ \ \ \ \times\left(
\begin{array}{l}
 \rho(t,\vec x)G(s,\vec x-\vec y)\rho(t-s,\vec y) \\
+\rho(t-s,\vec x)G(-s,\vec x-\vec y)\rho(t,\vec y)
\end{array}
\right).\nonumber
\end{eqnarray}
The final expression is obtained by noting that the information of the initial time $t_0$ will become irrelevant after (a few times) the finite correlation time of gluons.
The gluon correlation time is much shorter than the dynamical time scales of the heavy quark systems, such as the relaxation time.
The former is $\sim 1/gT$ or shorter ($\sim 1/T$) while the latter is $\sim 1/g^2T \gg 1/gT$ for decoherence and color diffusion (the kinetic relaxation time is much longer, $\sim M/g^4T^2$).
This condition corresponds to $\tau_{\rm E}\ll\tau_{\rm R}$ for the quantum Brownian motion in Table \ref{tab:oqs_regimes}.
Using the symmetry of gluon two-point functions, such as $G^{\rm F(\tilde F)}_{ab,00}(-s,-\vec r)=G^{\rm F(\tilde F)}_{ba,00}(s,\vec r)$ and $G^>_{ab,00}(-s,-\vec r)=G^<_{ba,00}(s,\vec r)$, we obtain
\begin{eqnarray}
\label{eq:IF_latetime}
iS_{\rm IF}\left[\rho_1,\rho_2\right]
&\simeq&-g^2\int^{\infty}_{t_0} dt \int d^3x d^3y\int_0^{\infty} ds
\left(
\rho^{a}_1,\ \rho^{a}_2
\right)_{(t,\vec x)}\nonumber\\
&\times&\left[
\begin{array}{cc}
G^{\rm F}_{ab,00}  & -G^<_{ab,00} \\
-G^>_{ab,00}       & G^{\rm \tilde F}_{ab,00}
\end{array}
\right]_{(s,\vec x-\vec y)}
\left(
\begin{array}{c}
\rho^{b}_1\\
\rho^{b}_2
\end{array}
\right)_{(t-s,\vec y)}. \nonumber\\
\end{eqnarray}
This influence functional is still nonlocal in time.
By the Markov approximation that will be made shortly, the influence functional becomes local in time and thus Markovian master equations will be obtained.

\subsubsection{Coarse graining in time}
When the intrinsic (not dynamical) time scale of the heavy quark color density is long compared to the gluon correlation time, we can perform a coarse graining in time as is commonly done in the derivation of quantum Brownian motion.
This corresponds to the condition $\tau_{\rm E}\ll\tau_{\rm S}$ in Table \ref{tab:oqs_regimes}.
The time scales of the heavy quark color density are $v/\dot v\sim \infty$ for single heavy quark kinetics and $v/\dot v \sim 1/M\alpha^2$ for relative motion in a quarkonium (assuming the Coulomb bound states), while the gluon correlation time is $\sim 1/gT$ or shorter ($\sim 1/T$).
Therefore, if $M\alpha^2\ll gT$ is satisfied, the quantum Brownian motion approach can also be applicable to a quarkonium.
In this case, we can neglect the effect of acceleration by the potential force during a scattering.

Schematically, the coupling of the heavy quark color densities at different times is approximated by truncating the following expansions:
\begin{eqnarray}
&&\int_0^{\infty}ds \ G(s)\rho(t)\rho(t-s)\nonumber\\
&&=\sum_{n=0}^{\infty}
\frac{1}{n!}[\rho(t)(i\partial_t)^n\rho(t)]
\int_{-\infty}^{\infty}\frac{d\omega}{2\pi i}\frac{\partial_{\omega}^n\tilde G(\omega)}{\omega - i\epsilon},
\end{eqnarray}
where $\tilde G(\omega)=\int dt e^{i\omega t} G(t,\vec r)$ and $\epsilon>0$.
The truncation corresponds to focusing on the long time behavior of the heavy quark color density $\rho(t)$.
In our case, we keep the terms with $n\leq2$, which corresponds to neglecting the effect of acceleration (after partial integration in time for $n=2$).
Since $\dot \rho^a=-\vec \nabla \cdot\vec j^a\sim \vec v\cdot\vec\nabla \rho^a$,  it formally takes a form of velocity expansion.

Using $G^{\rm F(\tilde F)}_{ab,00}(s,\vec r)=G^{>(<)}_{ab,00}(s,\vec r)=-\frac{i}{2}(\frac{i}{2})G^{\rm R}_{ab,00}(s,\vec r)+\frac{1}{2}G^{\rm S}_{ab,00}(s,\vec r)$ for $s>0$, Eq.~\eqref{eq:IF_latetime} can be expressed with $G^{\rm R}_{ab,00}(s,\vec r)$ and $G^{\rm S}_{ab,00}(s,\vec r)$.
For the couplings with the retarded propagator, the analytic structure of $\tilde G^{\rm R}_{ab,00}(\omega,\vec r)$ in the complex $\omega$ plane leads to
\begin{eqnarray}
\label{eq:coarse_retarded}
&&\int_0^{\infty}ds \ G^{\rm R}_{ab,00}(s,\vec x-\vec y)
\rho^a(t,\vec x)\rho^b(t-s,\vec y)\\
&&\simeq\sum_{n=0,1,2}
\frac{1}{n!}[\rho^a(t,\vec x)(i\partial_t)^n\rho^b(t,\vec y)]
\partial_{\omega}^n\tilde G^{\rm R}_{ab,00}(0,\vec x-\vec y).\nonumber
\end{eqnarray}
For the couplings with the symmetrized correlation function, we obtain 
\begin{eqnarray}
\label{eq:coarse_symmetrized}
&&\int_0^{\infty}ds \ G^{\rm S}_{ab,00}(s,\vec x-\vec y)
\rho^a(t,\vec x)\rho^b(t-s,\vec y)\\
&&\simeq
\frac{1}{2}\sum_{n=0,2}
\frac{1}{n!}[\rho^a(t,\vec x)(i\partial_t)^n\rho^b(t,\vec y)]
\partial_{\omega}^n\tilde G^{\rm S}_{ab,00}(0,\vec x-\vec y)\nonumber\\
&& \ \ \ +[\rho^a(t,\vec x)i\partial_t\rho^b(t,\vec y)]
\int_{-\infty}^{\infty} \frac{d\omega}{2\pi i}\frac{1}{\omega}
\partial_{\omega}\tilde G^{\rm S}_{ab,00}(\omega,\vec x-\vec y)\nonumber\\
&&\simeq
\frac{1}{2}\sum_{n=0,2}
\frac{1}{n!}[\rho^a(t,\vec x)(i\partial_t)^n\rho^b(t,\vec y)]
\partial_{\omega}^n\tilde G^{\rm S}_{ab,00}(0,\vec x-\vec y),\nonumber
\end{eqnarray}
using the fact that $\tilde G^{\rm S}_{ab,00}(\omega,\vec r)=\coth(\omega/2T)\sigma_{ab,00}(\omega,\vec r)$ is an even function of $\omega$.
Here, we drop the indices of the time contour in $\rho^a$.
By approximating the spectral function by an Ohmic one $\sigma_{ab,00}(\omega,\vec r)\sim \gamma_{ab}(\vec r)\omega$ with a cutoff at $|\omega|=\Omega \ll gT$ or $T$ to ignore the memory effect of gluons, the integral in the third line of Eq.~\eqref{eq:coarse_symmetrized} turns out to be $\propto \Omega/T \ll 1$ and thus can be ignored.
%\footnote{
%Another way to approximate the spectral function is to assume an Ohmic spectral function with a Lorentz-Drude cutoff $\sigma_{ab,00}(\omega,\vec r)\sim \gamma_{ab}(\vec r)\omega\Omega^2/(\omega^2 + \Omega^2)$.
%In this case, the argument in the text does not hold, but the third line of Eq.~\eqref{eq:coarse_symmetrized} is evaluated to be $(\dot\rho/\rho)/\Omega\sim (\Omega Ml_{\rm fluct}^2)^{-1}\sim  gT/M\ll 1$ (without a complex phase) times smaller than the second line for $n=0$ and thus can be neglected ($l_{\rm fluct}$ is defined later in the text).
%A similar argument is found in the derivation of the master equation for quantum Brownian motion \cite{BrePetText}.
%Note that with this spectral function $A(\vec r)\simeq D(\vec r)/6T^2$ in Eq.~\eqref{eq:new} corresponds to picking up minimal terms of $\mathcal O(g^2v^2)$ in the influence functional.
%}

The choice of $t$ matters because, if we took it to be $t=(x^0+y^0)/2$, we would have integrals of the form $\int_{t_0}^{\infty}dt\int_{-\infty}^{\infty}ds G(s)\rho(t)\rho(t-s)$.
Then $G^{\rm A}_{ab,00}(s,\vec r)$ would also contribute in $G^{\rm F(\tilde F)}_{ab,00}(s,\vec r)$ for $s<0$ and cancel the diagonal parts of Eq.~\eqref{eq:IF_diss} or \eqref{eq:IF_diss2} in the final result.
The reason why we have to take $t={\rm max}(x^0,y^0)$ will become clear when we discuss how to obtain the functional master equation.

\subsubsection{Influence functional in the Markov limit}
Let us define the following three functions to parametrize the influence functional:
\begin{eqnarray}
\label{eq:potential}
V(\vec r)\delta_{ab}&\equiv& -g^2 {\rm Re}\tilde G^{\rm R}_{ab,00}(0,\vec r),\\
\label{eq:fluctuation-dissipation}
D(\vec r)\delta_{ab}&\equiv& -g^2T\frac{\partial}{\partial\omega}
  \sigma_{ab,00}(0,\vec r),\\
\label{eq:new}
A(\vec r)\delta_{ab}&\equiv& -g^2\left(\
\frac{1}{6T}\frac{\partial}{\partial \omega}
+\frac{T}{3}\frac{\partial^3}{\partial\omega^3}
\right)
\sigma_{ab,00}(0,\vec r)\nonumber\\
&\simeq& -\frac{g^2}{6T}\frac{\partial}{\partial \omega}
\sigma_{ab,00}(0,\vec r),
\end{eqnarray}
where the Ohmic spectral function for $\sigma_{00,ab}(\omega,\vec r)$ is assumed as before to obtain $A(\vec r)\simeq D(\vec r)/6T^2$.
Explicit forms of ${\rm Re}\tilde G^{\rm R}_{ab,00}(0,\vec r)$ and $\frac{\partial}{\partial\omega}\sigma_{ab,00}(0,\vec r)$ at typical distance $r\sim 1/gT$ are given in Appendix \ref{sec:2point_functions}, using the hard thermal loop (HTL) resummed perturbation theory.
Using these functions, the influence functional in the Markov limit is given by four terms:
\footnote{
To obtain the influence functional \eqref{eq:IF_Markov}, we need to change the variables $\vec x\leftrightarrow\vec y$ in the integral to cancel some terms.
However, this apparently trivial operation is possible only for terms with $\rho_1^a(\vec x)(i\partial_t)^n\rho_2^a(\vec y)$ or $\rho_2^a(\vec x)(i\partial_t)^n\rho_1^a(\vec y)$.
For terms with $\rho_1^a(\vec x)(i\partial_t)^n\rho_1^a(\vec y)$ or $\rho_2^a(\vec x)(i\partial_t)^n\rho_2^a(\vec y)$, the variables $\vec x$ and $\vec y$ indicate the original order in time, which is essential when deriving the functional differential equation.
Such a problem does not occur for terms with $\rho_1^a(\vec x)(i\partial_t)^n\rho_2^a(\vec y)$ or $\rho_2^a(\vec x)(i\partial_t)^n\rho_1^a(\vec y)$.
}
\begin{eqnarray}
\label{eq:IF_Markov}
S_{\rm IF}&=&S_{\rm pot} + S_{\rm fluct} + S_{\rm diss} + S_{\rm L} + \cdots,\\
\label{eq:IF_pot}
iS_{\rm pot}&=&-\frac{i}{2}\int_{t_0} dt\int d^3xd^3yV(\vec x-\vec y) \\
&& \ \times\left(
\rho^{a}_1,\ \rho^{a}_2
\right)_{(t,\vec x)}
\left[
\begin{array}{cc}
1 & 0 \\
0 & -1
\end{array}
\right]
\left(
\begin{array}{c}
\rho^{a}_1\\
\rho^{a}_2
\end{array}
\right)_{(t,\vec y)}, \nonumber \\
\label{eq:IF_fluct}
iS_{\rm fluct}&=&-\frac{1}{2}\int_{t_0} dt\int d^3xd^3yD(\vec x-\vec y)\\
&& \ \times\left(
\rho^{a}_1,\ \rho^{a}_2
\right)_{(t,\vec x)}
\left[
\begin{array}{cc}
-1 & 1 \\
1 & -1
\end{array}
\right]
\left(
\begin{array}{c}
\rho^{a}_1\\
\rho^{a}_2
\end{array}
\right)_{(t,\vec y)}, \nonumber
\end{eqnarray}
\begin{eqnarray}
\label{eq:IF_diss}
iS_{\rm diss}&=&-\frac{i}{4T}\int_{t_0} dt\int d^3xd^3yD(\vec x-\vec y)\\
&& \ \times\left(
\rho^{a}_1,\ \rho^{a}_2
\right)_{(t,\vec x)}
\left[
\begin{array}{cc}
-1 & -1 \\
1  &  1
\end{array}
\right]
\left(
\begin{array}{c}
\dot\rho^{a}_1\\
\dot\rho^{a}_2
\end{array}
\right)_{(t,\vec y)}, \nonumber\\
\label{eq:IF_acc}
iS_{\rm L}&=&-\frac{1}{4}\int_{t_0} dt\int d^3xd^3yA(\vec x-\vec y)\\
&& \ \times\left(
\dot\rho^{a}_1,\ \dot\rho^{a}_2
\right)_{(t,\vec x)}
\left[
\begin{array}{cc}
-1 & 1 \\
1 & -1
\end{array}
\right]
\left(
\begin{array}{c}
\dot\rho^{a}_1\\
\dot\rho^{a}_2
\end{array}
\right)_{(t,\vec y)}. \nonumber
\end{eqnarray}
Each term has physical meanings: 
$S_{\rm pot}$ gives a potential between two heavy quarks,
$S_{\rm fluct}$ accounts for thermal fluctuations,
$S_{\rm diss}$ gives rise to dissipative dynamics such as the drag force, 
and $S_{\rm L}$, which is proportional to $(\dot\rho_{1}-\dot\rho_{2})(\dot\rho_{1}-\dot\rho_{2})$, is a new term first introduced in this paper and makes an essential contribution to render the Lindblad-form master equations.
This is analogous to the situation in the quantum Brownian motion \cite{Caldeira:1982iu}, where it is necessary to include the $(\dot x_1-\dot x_2)^2$ term in the influence functional in order to obtain the Lindblad-form master equation \cite{Diosi:1993}.

The counting in the perturbative and velocity expansion is $S_{\rm pot}, S_{\rm fluct}\sim g^2v^0$, $S_{\rm diss}\sim g^2v$, and $S_{\rm L}\sim g^2v^2$.
In the counting, the order of $V(\vec r), D(\vec r), A(\vec r)$  is loosely counted as $\mathcal O(g^2)$ for all $\vec r$ and similarly for their derivatives.
We keep it loose unless it is worth making it more precise.
This is because our description in the regime of quantum Brownian motion is not confined to particular states (such as 1S and 2S states) and thus the spatial size of the wave function is not necessarily determined uniquely.
In $S_{\rm pot}$, we ignore a term $\propto (\dot\rho_1\dot\rho_1-\dot\rho_2\dot\rho_2)$ because it would just give an $\mathcal O(g^2v^2)$ correction to the potential.
To be strict, this is not consistent with the velocity expansion, but we keep $S_{\rm L}$ in order to obtain the master equations in the Lindblad form.

It should also be remarked that we also implicitly rely on the perturbative expansion in the procedure of coarse graining in time.
The couplings in $S_{\rm IF}$ is originally nonlocal in time.
Because of the coarse graining, the couplings are approximated to be local.
This approximation corresponds to the ladder approximation in the Bethe-Salpeter equation.
The overlap of two interactions is thus neglected, which would yield a cross-ladder contribution of higher order in $g$.

As discussed before, one can use the free equations of motion for $\dot\rho_1$ and $\dot\rho_2$ in the coarse graining and hence $S_{\rm diss}$ and $S_{\rm L}$ become
\begin{eqnarray}
\label{eq:IF_diss2}
iS_{\rm diss}&=&\frac{i}{4T}\int_{t_0} dt\int d^3xd^3y\vec\nabla_xD(\vec x-\vec y)\\
&& \ \cdot\left(
\rho^{a}_1,\ \rho^{a}_2
\right)_{(t,\vec x)}
\left[
\begin{array}{cc}
-1 & -1 \\
1  &  1
\end{array}
\right]
\left(
\begin{array}{c}
\vec j^{a}_1\\
\vec j^{a}_2
\end{array}
\right)_{(t,\vec y)}, \nonumber\\
\label{eq:IF_acc2}
iS_{\rm L}&=&\frac{1}{4}\int_{t_0} dt\int d^3xd^3y
\nabla_x^k\nabla_x^lA(\vec x-\vec y)\\
&& \ \times\left(
j^{a}_1,\ j^{a}_2
\right)^k_{(t,\vec x)}
\left[
\begin{array}{cc}
-1 & 1 \\
1 & -1
\end{array}
\right]
\left(
\begin{array}{c}
j^{a}_1\\
j^{a}_2
\end{array}
\right)^l_{(t,\vec y)}. \nonumber
\end{eqnarray}
Equations \eqref{eq:IF_Markov}-\eqref{eq:IF_fluct} and \eqref{eq:IF_diss2}-\eqref{eq:IF_acc2} constitute the influence functional in the leading orders in perturbative and velocity expansions up to the order of $\mathcal O(g^2v^0,g^2v)$ [and some terms of order $\mathcal O(g^2v^2)$] in the Markov limit.
Note that here we only consider the color density interaction in the heavy quark sector, which remains in the $1/c\to 0$ limit.

\subsection{Functional master equations}
Here, we review how to obtain the renormalized effective Hamiltonian described in Ref.~\cite{Akamatsu:2012vt}.
The total action $S_{\rm CTP}=\int_{t_0} dt L_{\rm CTP}$ on the closed-time path is given by adding nonrelativistic kinetic terms for $\psi_{1,2}=(Q,Q^{\dagger}_c)_{1,2}$, where $Q_{(c)}$s are Pauli spinors for a heavy (anti)quark:
\begin{eqnarray}
S_{\rm CTP}[\psi_1,\psi_2]&=&S_{\rm kin}[\psi_1]-S_{\rm kin}[\psi_2]+S_{\rm IF}[\rho_1,\rho_2],\\
S_{\rm kin}[\psi] &=&
\int_{t_0}d^4x Q^{\dagger}\left(i\partial_0-M+\frac{\nabla^2}{2M}\right)Q\\
&&+\int_{t_0}d^4x Q_c\left(i\partial_0+M-\frac{\nabla^2}{2M}\right)Q^{\dagger}_c. \nonumber
\end{eqnarray}
Since the partition function is
\begin{eqnarray}
Z[\eta_1,\eta_2]&=&\int {\mathcal D}[\psi_1,\psi_2]
\langle \psi_1^{\dagger} |{\bm\rho}_{\rm S}|\psi_2 \rangle_{t_0}
e^{iS_{\rm CTP}}\\
&& \times \ e^{
i\int_{t_0} d^4x \left(
\eta_1^{\dagger}\psi_1+\psi_1^{\dagger}\eta_1
-\eta_2^{\dagger}\psi_2-\psi_2^{\dagger}\eta_2
\right)},\nonumber
\end{eqnarray}
where ${\bm\rho}_{\rm S}$ is the arbitrary initial density matrix in the heavy quark Fock space,
the reduced density matrix at later time $t'>t_0$ is given by
\begin{eqnarray}
&&\langle {\psi'}_1^{\dagger} |{\bm\rho}_{\rm S}(t')|\psi'_2 \rangle \nonumber\\
&&=\int^{{\psi'}_1^{\dagger},\psi'_2} {\mathcal D}[\psi_1,\psi_2]
\langle \psi_1^{\dagger} |{\bm\rho}_{\rm S}|\psi_2 \rangle_{t_0}
e^{i\int_{t0}^{t'} dt L_{\rm CTP}},
\end{eqnarray} 
with boundary conditions $\psi_1^{\dagger}(t')={\psi'}_1^{\dagger}$ and $\psi_2(t')=\psi'_2$.
Note that time integration is limited to $t<t'$.
This is why we must choose $t={\rm max}(x^0,y^0)$ in the previous section.

The time-evolution equation for $\rho_{\rm S}\left[t,\psi_1^{\dagger},\psi_2\right]\equiv\langle \psi_1^{\dagger} |{\bm\rho}_{\rm S}(t)|\psi_2 \rangle$ is given by an analogy with the Schr\"odinger equation.
\begin{enumerate}
\item
Derive the {\it Hamiltonian} $\bm H_{\rm CTP}\left[\bm \psi_1^{\dagger},\bm\psi_1,\bm\psi_2^{\dagger},\bm\psi_2\right]$ corresponding to the Lagrangian $L_{\rm CTP}$ by the Legendre transformation.
\item
Obtain functional representation of $\bm H_{\rm CTP}$ by the following replacement:
\begin{eqnarray}
&&\bm H_{\rm CTP}\left[\bm\psi_1^{\dagger},\bm\psi_1,\bm\psi_2^{\dagger},\bm\psi_2\right]\nonumber\\
&&\to
H_{\rm CTP}\left[\psi_1^{\dagger},\frac{\delta}{\delta\psi_1^{\dagger}},-\frac{\delta}{\delta\psi_2},\psi_2\right].
\end{eqnarray}
\item
The functional master equation is obtained as
\begin{eqnarray}
&&i\frac{\partial}{\partial t}\rho_{\rm S}\left[t,\psi_1^{\dagger},\psi_2\right]\nonumber \\
&&=H_{\rm CTP}\left[\psi_1^{\dagger},\frac{\delta}{\delta\psi_1^{\dagger}},-\frac{\delta}{\delta\psi_2},\psi_2\right]\rho_{\rm S}\left[t,\psi_1^{\dagger},\psi_2\right].
\end{eqnarray}
\end{enumerate}
In the first step, we must take care of the order of operators, which must be ordered by time.
For example, in the fermion bilinear in $\rho^a_{1,2}(t,\vec x)$ and in the kinetic terms, time is assigned as $\psi_1^{\dagger}(t+\epsilon),\psi_1(t-\epsilon)$ and $\psi_2^{\dagger}(t-\epsilon),\psi_2(t+\epsilon)$ with $\epsilon>0$.
Also as is clear from Eq.~\eqref{eq:IF_latetime}, the time for $\rho^a_{1,2}(t,\vec x)$ is later than that for $\rho^a_{1,2}(t,\vec y)$ in Eq.~\eqref{eq:IF_Markov}.
Use of new variables $\tilde\psi_2=\psi_2^{\dagger}, {\tilde\psi}^{\dagger}_2=\psi_2$ will make the fields on the 1 and 2 axes look symmetric.

Since we are interested in systems with a few heavy quarks in the QGP, coherent states $\langle Q_{1(c)}^*|$ and $|{\tilde Q}^*_{2(c)}\rangle$ defined as
\begin{eqnarray}
\label{eq:coherent1}
\langle Q_{1(c)}^*|&=&\langle\Omega| e^{-\int d^3x\left(
\bm Q(\vec x)Q_1^*(\vec x) + \bm Q_c(\vec x)Q_{1c}^*(\vec x)\right)},\\
\label{eq:coherent2}
|\tilde Q_{2(c)}^*\rangle&=&e^{-\int d^3x\left(
{\tilde Q}_2^*(\vec x)\bm Q^{\dagger}(\vec x) + {\tilde Q}_{2c}^*(\vec x)\bm Q^{\dagger}_c(\vec x)\right)}
|\Omega\rangle
\end{eqnarray}
are more convenient to express $\bm\rho_{\rm S}(t)$.
Here, $|\Omega\rangle$ is the vacuum state that satisfies $\bm Q_{(c)}|\Omega\rangle = 0$.
This amounts to changing the variables for functional differentiation,
\begin{eqnarray}
&&\bm H_{\rm CTP}\left[
\bm Q_{1(c)}^{\dagger},\bm Q_{1(c)},
\tilde{\bm Q}_{2(c)}^{\dagger},\bm \tilde{\bm Q}_{2(c)}
\right]\nonumber\\
&&\to H_{\rm CTP}\left[
Q_{1(c)}^*,\frac{\delta}{\delta Q_{1(c)}^*},
\tilde Q_{2(c)}^*,-\frac{\delta}{\delta \tilde Q_{2(c)}^*}
\right],
\end{eqnarray}
and the functional master equation is given by
\begin{eqnarray}
\label{eq:func_master}
&&i\frac{\partial}{\partial t}\rho_{\rm S}\left[t,Q_{1(c)}^*,{\tilde Q}_{2(c)}^*\right]\\
&&=H_{\rm CTP}\left[
Q_{1(c)}^*,\frac{\delta}{\delta Q_{1(c)}^*},
\tilde Q_{2(c)}^*,-\frac{\delta}{\delta \tilde Q_{2(c)}^*}
\right]\nonumber \\
&& \ \ \times\rho_{\rm S}\left[t,Q_{1(c)}^*,{\tilde Q}_{2(c)}^*\right].\nonumber
\end{eqnarray}
In general, the time-ordered product does not give an operator $H_{\rm CTP}$ in such a way that all the differentiation is moved on the right.
Therefore, in the course of doing so after deriving the time-ordered $H_{\rm CTP}$, we need to subtract divergent contributions in the vacuum, e.g., Coulomb potential at the origin $V(\vec 0)$ in the self energy, by introducing counterterms.

\subsection{From fields to particles}
The functional master equation can generate master equations for systems with an arbitrary finite number of heavy quarks in the QGP.
Since the coherent states act as a generating functional for heavy quarks as in Eqs.~\eqref{eq:coherent1}-\eqref{eq:coherent2}, the reduced density matrices are given by functionally differentiating $\rho_{\rm S}\left[t,Q_{1(c)}^*,{\tilde Q}_{2(c)}^*\right]$.
For example, the reduced density matrix of a single heavy quark system in the QGP is obtained by
\begin{eqnarray}
\label{eq:particle_density_matrix}
\rho^{ij}_{Q}(t,\vec x,\vec y)
&=&\langle \vec x,i|\bm \rho_Q(t)|\vec y,j\rangle\\
&=& \langle \Omega| \bm Q^i(\vec x) \bm\rho_{\rm S}(t)\bm Q^{j\dagger}(\vec y)|\Omega\rangle \nonumber\\
&=&-\frac{\delta}{\delta Q^{i*}_1(\vec x)}\frac{\delta}{\delta \tilde Q_2^{j*}(\vec y)}
\rho_{\rm S}\left[t,Q^*_{1(c)},\tilde Q_{2(c)}^*\right]\Big |_{Q^*=0}.\nonumber
\end{eqnarray}
Therefore, in order to obtain the master equations for heavy quark reduced density matrices, we just need to perform appropriate functional differentiations on both sides of the functional master equation \eqref{eq:func_master} and switch off the source $Q_{1(c)}^*=\tilde Q_{2(c)}^*=0$.
In Appendix \ref{sec:illustration}, we illustrate how a term $-\frac{1}{2}\int_{t^0}dt\int d^3xd^3y D(\vec x-\vec y)\rho^a_1(t,\vec x)\rho^a_2(t,\vec y)\in iS_{\rm fluct}$ contributes in the master equation for a single heavy quark as an example.

Similarly, the forward propagators of heavy quarks are given by differentiating only with $Q_{1(c)}^*$ fields, up to the correction of order $\mathcal O(e^{-M/T})\ll 1$.
For example, the one-body forward propagator is obtained by
\begin{eqnarray}
G^>_{Q,i}(t,\vec x) &=& 
\frac{
{\rm Tr}\left[e^{-{\bm H_{\rm QCD}}/T}\bm Q^i(t,\vec x)\bm Q^{j\dagger}(t_0,\vec y)\right]
}{
{\rm Tr}\left[e^{-{\bm H_{\rm QCD}}/T}\right]
}\nonumber\\
&\simeq&\frac{\delta}{\delta Q^{i*}_1(\vec x)}\rho_{\rm S}\left[t,Q^*_{1(c)},\tilde Q_{2(c)}^*\right]\Big |_{Q^*=0}.
\end{eqnarray}
Therefore, the time-evolution equations for forward propagators are also derived from the functional master equation \eqref{eq:func_master} by performing appropriate functional differentiations with $Q^*_{1(c)}$ fields.
In Appendix \ref{sec:forward_propagator}, we show the time-evolution equation for the forward propagator of a quarkonium, for which the leading correction in the velocity expansion is found to be $\mathcal O(v)$.
The $\mathcal O(v)$ term couples relative position and momentum of a heavy quark-antiquark pair in an intriguing way.
It may be necessary to take into account the $\mathcal O(v)$ term when one computes the vector current spectral function using the complex potential.

\section{Master equations in the Lindblad form}
\label{sec:master_equations}
In this section, we derive master equations for a single heavy quark and a quarkonium in the QGP.
We show that each master equation can be written in the Lindblad form
\begin{eqnarray}
\label{eq:Lindblad2}
\frac{d}{dt}\bm\rho_{\rm S}(t)&=&-i[\bm H,\bm\rho_{\rm S}] \\
&+&\sum_{i=1}^{N}\gamma_i\left(
\bm L_i \bm\rho_{\rm S} \bm L^{\dagger}_i
-\frac{1}{2} \bm L_i^{\dagger} \bm L_i \bm\rho_{\rm S}
-\frac{1}{2}\bm\rho_{\rm S} \bm L_i^{\dagger} \bm L_i
\right),\nonumber
\end{eqnarray}
with $\bm H^{\dagger}=\bm H$ and $\gamma_i>0$ for ${\rm S}=Q$ and $QQ_c$.
This is equivalent to showing the master equations preserving complete positivity of the reduced density matrices.

\subsection{Single heavy quark master equations}
\subsubsection{Full master equation}
By following the procedures outlined in the previous section, the master equation for the reduced density matrix of a single heavy quark $\hat\rho_Q(t,\vec x,\vec y)$ [$N_{\rm c}\otimes N^*_{\rm c}$ representation of the color ${\rm SU}(N_{\rm c})$ group] is obtained as
\begin{eqnarray}
\label{eq:HQ_master_full}
&&\frac{\partial}{\partial t}\hat\rho_{Q}(t,\vec x,\vec y)
=i\frac{\vec\nabla_x^2-\vec\nabla_y^2}{2M}\hat\rho_{Q}(t,\vec x,\vec y)\nonumber\\
&& \ \ \ + \ F_1(\vec x-\vec y)t^a\hat\rho_{Q}(t,\vec x,\vec y)t^a
-C_{\rm F}F_1(\vec 0)\hat\rho_{Q}(t,\vec x,\vec y)\nonumber \\
&& \ \ \ + \ \vec F_2(\vec x-\vec y)\cdot (\vec \nabla_x -\vec \nabla_y)t^a\hat\rho_{Q}(t,\vec x,\vec y)t^a\nonumber\\
&& \ \ \ + \ F^{ij}_3(\vec x-\vec y)\nabla^i_x\nabla^j_y 
t^a\hat\rho_{Q}(t,\vec x,\vec y)t^a\nonumber \\
&& \ \ \ + \ C_{\rm F}F^{ii}_3(\vec 0)\frac{\vec\nabla^2_x+\vec\nabla^2_y}{6}
\hat\rho_{Q}(t,\vec x,\vec y),
\end{eqnarray}
where $C_{\rm F}=(N_{\rm c}^2-1)/2N_{\rm c}$ and
\begin{eqnarray}
F_1(\vec r)&=& -\left(D(\vec r)+\frac{\vec\nabla^2 D(\vec r)}{4MT}+\frac{(\vec\nabla^2)^2 A(\vec r)}{8M^2}\right),\\
\vec F_2(\vec r)&=& -\vec\nabla\left(\frac{D(\vec r)}{4MT}
+\frac{\vec\nabla^2 A(\vec r)}{4M^2}\right),\\
F^{ij}_3(\vec r)&=& \nabla^i\nabla^j\left(\frac{ A(\vec r)}{2M^2}\right).
\end{eqnarray}
In the master equation \eqref{eq:HQ_master_full}, there are terms with different factors of $1/M$.
As can be understood from the factors of $1/M$ in the influence functional, $(1/M)^0$ terms come from $S_{\rm fluct}$, $1/M$ from $S_{\rm diss}$, and $(1/M)^2$ from $S_{\rm L}$.
All the terms in the master equation can be evaluated by the orders of perturbation $g$, velocity $v$, and $T/M(\equiv\delta)$.
Here velocity comes from $\nabla\sim Mv$ acting on the reduced density matrix.
For example, $F_1(\vec x-\vec y)t^a\hat\rho_Q(t,\vec x,\vec y)t^a$ consists of $\mathcal O(g^2v^0\delta^0),\mathcal O(g^2v^0\delta)$, and $\mathcal O(g^2v^0\delta^2)$ terms.
Note that terms of $\mathcal{O}(v^n)$ in the influence functional $S_{\rm IF}$ yield terms of $\mathcal{O}(v^l\delta^m)$ with $n=l+m \ (l,m\geq 0)$ in the master equation.
For instance, $S_{\rm diss}\sim g^2v$ produces $\mathcal O(g^2v^0\delta)$ (the second term in $F_1$) as well as $\mathcal O(g^2v\delta^0)$ terms (the first term in $F_2$).
This happens because some of the derivatives in $\vec j^a$ in Eqs.~\eqref{eq:IF_diss2}-\eqref{eq:IF_acc2} act on $D(\vec x-\vec y)$ or $A(\vec x-\vec y)$, not on $Q^*$, in the course of deriving the functional master equation \eqref{eq:func_master}.
Owing to this mismatch in the counting, an approximation to the master equation and that to the influence functional may not be consistent with each other.
In this respect, the influence functional is more fundamental than the master equations.
Therefore, we always make such approximations to the master equations that can be derived from approximated influence functionals.

The master equation \eqref{eq:HQ_master_full} can be written in the Lindblad form Eq.~\eqref{eq:Lindblad2} with $\bm H = \vec{\bm p}^2/2M$.
The label is $i=(\vec k,a,\alpha)$, where $\vec k$ is the wave number in a box with volume $L^3$, $a$ is the label for color matrix $t^a$, and $\alpha=1,2$ is introduced for classification.
The Lindblad operators $\bm L^{\alpha}_{\vec k a}$ and coefficients $\gamma^{\alpha}_{\vec k a}$ are
\begin{eqnarray}
\label{eq:HQ_Lindblad_a1}
&&\left\{
\begin{array}{l}
\bm{L}^{\alpha=1}_{\vec k a}=
e^{i\vec k\cdot\vec {\bm x}/2}
\left(1-\frac{\vec k\cdot\vec{\bm p}}{4MT}\right)
e^{i\vec k\cdot\vec {\bm x}/2}\bm t^a, \\
\gamma^{\alpha=1}_{\vec k a}=-\frac{\tilde D(\vec k)}{L^3} > 0,
\end{array}\right.\\
\label{eq:HQ_Lindblad_a2}
&&\left\{
\begin{array}{l}
\bm{L}^{\alpha=2}_{\vec k a}=
e^{i\vec k\cdot\vec {\bm x}/2}
\left(\frac{\vec k\cdot\vec{\bm p}}{4MT}\right)
e^{i\vec k\cdot\vec {\bm x}/2}\bm t^a, \\
\gamma^{\alpha=2}_{\vec k a}=-\frac{1}{L^3}
\left(8T^2\tilde A(\vec k)-\tilde D(\vec k)\right) > 0.
\end{array}\right.
\end{eqnarray}
Here, $\tilde D(\vec k)=\int d^3r e^{-i\vec k\cdot\vec r}D(\vec r)$ and similarly for $\tilde A(\vec k)$.
Without the term $\tilde A(\vec k)$, or $S_{\rm L}$ in the influence functional, the coefficient $\gamma^{\alpha=2}_{\vec ka}$ is negative and the master equation cannot be in the Lindblad form.
Therefore, keeping $S_{\rm L}$ together with $S_{\rm diss}$ in the influence functional is essential in obtaining the Lindblad-form master equation.

By tracing out the color space dynamics $\bar \rho_Q(t,\vec x,\vec y)={\rm Tr}_{\rm color}\hat \rho_Q(t,\vec x,\vec y)=\rho_Q^{ii}(t,\vec x,\vec y)$, the master equation for $\bar \rho_Q(t,\vec x,\vec y)$ reads
\begin{eqnarray}
\label{eq:HQ_master_ctr}
&&\frac{\partial}{\partial t}\bar\rho_{Q}(t,\vec x,\vec y)
=i\frac{\vec\nabla_x^2-\vec\nabla_y^2}{2M}\bar\rho_{Q}(t,\vec x,\vec y)\nonumber\\
&& \ \ \ + \ C_{\rm F}\left(F_1(\vec x-\vec y)-F_1(\vec 0)\right)
\bar\rho_{Q}(t,\vec x,\vec y)\nonumber \\
&& \ \ \ + \ C_{\rm F}\vec F_2(\vec x-\vec y)\cdot (\vec \nabla_x -\vec \nabla_y)\bar\rho_{Q}(t,\vec x,\vec y)\nonumber\\
&& \ \ \ + \ C_{\rm F}\left(
\begin{array}{c}
F^{ij}_3(\vec x-\vec y)\nabla^i_x\nabla^j_y \\
+F^{ii}_3(\vec 0)\frac{\vec\nabla^2_x+\vec\nabla^2_y}{6}
\end{array}
\right)
\bar\rho_{Q}(t,\vec x,\vec y),
\end{eqnarray}
and the Lindblad operators are obtained by replacing $\bm t^a$s with $\bm 1$ in Eqs.~\eqref{eq:HQ_Lindblad_a1}-\eqref{eq:HQ_Lindblad_a2} and the coefficients are $C_{\rm F}$ times those in Eqs.~\eqref{eq:HQ_Lindblad_a1}-\eqref{eq:HQ_Lindblad_a2}.

So far, we have not assumed a typical size of heavy quark wave functions.
In the next sections, the full master equation is approximated according to the wave function size.
We derive effective quantum dynamics for localized wave packets and extended wave functions.
These effective dynamics are summarized in Table \ref{tab:conditions_HQMEs}.

\begin{table}[top]
\begin{tabular}{c|c|c}
\hline
 & Wave packet & Recoilless limit \\
\hline
Wave function size & $\Delta x\sim l_{\rm dB}\ll l_{\rm fluct}$ & $\Delta x \gg l_{\rm dB}$ \\
\hline
\raisebox{2mm}{Approximation of $S_{\rm IF}$} & 
\shortstack{$D(\vec r)\simeq D_0+D_2 \vec r^2/6$ \\$A(\vec r)\simeq A_0+A_2 \vec r^2/6$ } &
\raisebox{2mm}{$S_{\rm IF}\simeq S_{\rm fluct}$} \\
\hline
Physical process & Langevin dynamics & Decoherence \\
\hline
\end{tabular}
\caption{
Summary of the approximated master equations \eqref{eq:HQ_master_wp} and \eqref{eq:HQ_master_rcless} for the single heavy quark.
}
\label{tab:conditions_HQMEs}
\end{table}

\subsubsection{Master equation for wave packets}
Now let us assume that the heavy quark is kinetically thermalized and its wave function is localized compared to the length scale of functions $D(\vec r)$ and $A(\vec r)$.
Close to heavy quark kinetic equilibrium, the heavy quark wave function extends over the thermal de Broglie wavelength $l_{\rm dB}\sim1/\sqrt{MT}$.
The size of the wave function is characterized by the ``correlation length'' $|\vec x-\vec y|$ of $\bar\rho_Q(t,\vec x, \vec y)$.
Therefore, in the master equation \eqref{eq:HQ_master_ctr}, we can approximate $D(\vec r)$ and $A(\vec r)$ by
\begin{eqnarray}
\label{eq:expansion_wp1}
D(\vec r)&\simeq& D(\vec 0)+\frac{{\vec r}^2}{6}\vec\nabla^2 D(\vec 0)
\equiv D_0 + \frac{D_2}{6}\vec r^2,\\
\label{eq:expansion_wp2}
A(\vec r)&\simeq& A(\vec 0)+\frac{{\vec r}^2}{6}\vec\nabla^2 A(\vec 0)
\equiv A_0 + \frac{A_2}{6}\vec r^2,
\end{eqnarray}
which yields
\begin{eqnarray}
\label{eq:HQ_master_wp}
&&\frac{\partial}{\partial t}\bar\rho_{Q}(t,\vec x,\vec y)
=i\frac{\vec\nabla_x^2-\vec\nabla_y^2}{2M}\bar\rho_{Q}(t,\vec x,\vec y)\nonumber \\
&& \ \ \ - \frac{C_{\rm F}D_2}{6}\left(
(\vec x-\vec y)^2 + (\vec x-\vec y)\cdot\frac{\vec \nabla_x-\vec \nabla_y}{2MT}
\right)\bar\rho_{Q}(t,\vec x,\vec y)\nonumber\\
&& \ \ \ + \frac{C_{\rm F}A_2}{12M^2}\left(
\vec \nabla_x + \vec \nabla_y
\right)^2\bar\rho_{Q}(t,\vec x,\vec y).
\end{eqnarray}
As is clear from Eqs.~\eqref{eq:expansion_wp1}-\eqref{eq:expansion_wp2}, this approximation can be made at the level of the influence functional.

By means of the counting in $g$, $\delta=T/M$, and $v\sim \sqrt{T/M}=\delta^{1/2}$, we can also make the above argument more precise.
The thermal de Broglie wavelength of a heavy quark $l_{\rm dB}\sim 1/\sqrt{MT}=\delta^{1/2}/T$ is much smaller than the length scale $l_{\rm fluct}\sim 1/gT$ of $D(\vec r)$ and $A(\vec r)$.
The latter is defined so that for $|\vec r|\agt l_{\rm fluct}$, $D(\vec r),A(\vec r)\simeq 0$ holds.
Then Eqs.~\eqref{eq:expansion_wp1}-\eqref{eq:expansion_wp2} are evaluated as expansions up to $(l_{\rm dB}/l_{\rm fluct})^2\sim g^2\delta$.
The master equation is also expanded in terms of $g^2\delta$.
Keeping the terms up to $\mathcal O(g^2\delta)$ in this expansion yields Eq.~\eqref{eq:HQ_master_wp}.
Using $D_2\sim D_0/l_{\rm fluct}^2\sim g^4T^3$ and $A_2\sim A_0/l_{\rm fluct}^2\sim (D_0/T^2)/l_{\rm fluct}^2$, the time scale of the Langevin dynamics of Eq.~\eqref{eq:HQ_master_wp} is estimated to be $\sim M/g^4T^2$.
To be strict, there is a logarithmic correction $\sim M/[g^4\ln(1/g)T^2]$ because $D_2$, which is proportional to a momentum diffusion constant, receives as much contribution from hard scatterings as from soft scatterings.
See Appendix \ref{sec:2point_functions} for details.

The Lindblad operators and coefficients are labeled with $i=(l,\alpha)$, where $l=x,y,z$,
\begin{eqnarray}
&&\left\{
\begin{array}{l}
\bm L^{\alpha=1}_l = \left(\vec {\bm x} + \frac{i\vec {\bm p}}{4MT}\right)_l,\\
\gamma^{\alpha=1}_l=\frac{C_{\rm F}D_2}{3}>0,
\end{array}\right.\\
&&\left\{
\begin{array}{l}
\bm L^{\alpha=2}_l = \left(\frac{\vec {\bm p}}{M}\right)_l,\\
\gamma^{\alpha=2}_l=\frac{C_{\rm F}}{48T^2}(8T^2A_2-D_2)>0,
\end{array}\right.
\end{eqnarray}
and the Hamiltonian is
\begin{eqnarray}
\bm H=\frac{\vec{\bm p}^2}{2M}
+\frac{C_{\rm F}D_2}{12MT}
\frac{\{\vec{\bm x}, \vec{\bm p}\}}{2}.
\end{eqnarray}
Here, the number of the Lindblad operators is reduced to only six and the Hamiltonian contains a term which is time-reversal odd.
If we neglect $A_2$, the master equation \eqref{eq:HQ_master_wp} is the same as that of the Caldeira-Leggett model of quantum Brownian motion.
Note that, without $A_2$, the second coefficient becomes $\gamma^{\alpha=2}_l < 0$ and the master equation is no longer in the Lindblad form.
Thus again, we find that $S_{\rm L}$ makes an essential contribution in obtaining the Lindblad-form master equation.

\subsubsection{Master equation in the recoilless limit}
If one is interested in decoherence of a wave function at distant points, which takes place much faster than the momentum dissipation, one can approximate the full master equation \eqref{eq:HQ_master_full} by just keeping the kinetic term and terms from $S_{\rm fluct}$ in the influence functional:
\begin{eqnarray}
\label{eq:HQ_master_rcless}
&&\frac{\partial}{\partial t}\hat\rho_{Q}(t,\vec x,\vec y)
=i\frac{\vec\nabla_x^2-\vec\nabla_y^2}{2M}\hat\rho_{Q}(t,\vec x,\vec y)\\
&& \ \ \ - \ D(\vec x-\vec y)t^a\hat\rho_{Q}(t,\vec x,\vec y)t^a
+C_{\rm F}D(\vec 0)\hat\rho_{Q}(t,\vec x,\vec y).\nonumber
\end{eqnarray}
This is called the recoilless limit of the full master equation.
Note that $S_{\rm pot}$ has no contribution to the master equation of a single heavy quark.

Let us examine in more detail under which conditions the decoherence takes place rapidly compared to the momentum dissipation.
The condition for the distance $\Delta x$ is $|F_1(\Delta \vec x)-F_1(\vec 0)|\gg |\vec F_2(\Delta \vec x)|Mv$ or $|D(\Delta \vec x)-D(\vec 0)|\gg |\vec\nabla D(\Delta \vec x)|v/4T$.
At large enough distance $\Delta x\agt l_{\rm fluct}$, where $D(\Delta \vec x)\simeq 0$ holds, the condition is satisfied.
At shorter distance $\Delta x\ll l_{\rm fluct}$, we can derive a condition $\Delta x \gg v/T$, that is $\Delta x \gg l_{\rm dB}$.
Therefore for $\Delta x\gg l_{\rm dB}$, the decoherence takes place more rapidly than momentum dissipation and the full master equation \eqref{eq:HQ_master_full} can be approximated by taking the recoilless limit.
The time scale depends on $\Delta x$: For $\Delta x\agt l_{\rm fluct}$ the time scale is $\sim 1/D(0)\sim 1/g^2T$ and for $l_{\rm fluct} \gg \Delta x \gg l_{\rm dB}$ the time scale is $\sim(1/D_2)/(\Delta x)^2\sim [g^4\ln(1/g)T^3(\Delta x)^2]^{-1}$.
Even if an initial wave function is coherent over $\Delta x\gg l_{\rm dB}$, its coherence is lost [$\hat \rho(t,\vec x,\vec y)\simeq 0$ for $|\vec x-\vec y|\simeq \Delta x$] through a few scatterings with medium particles.
Note that for heavy quarks to be kinetically thermalized, it requires many scatterings ($\propto M/T$) and thus takes a much longer time than decoherence.
Close to heavy quark kinetic equilibrium, the typical wave function is coherent only over $\Delta x\sim l_{\rm dB}$ and thus the master equation \eqref{eq:HQ_master_rcless} is not applicable there.

The master equation in the recoilless limit \eqref{eq:HQ_master_rcless} can be written in the Lindblad form.
The Lindblad operator is $\bm{L}_{\vec k a}=e^{i\vec k\cdot\vec {\bm x}}\bm t^a$ and the coefficient is $\gamma_{\vec k a}= - \tilde D(\vec k)/L^3>0$.
The Hamiltonian is $\bm H=\frac{\vec{\bm p}^2}{2M}$.
As mentioned before, the master equation \eqref{eq:HQ_master_rcless} is in the Lindblad form but cannot describe heavy quark kinetic equilibration.

The same approximation can be made to the color-traced master equation \eqref{eq:HQ_master_ctr}.
Or equivalently one can trace out the color space dynamics in the master equation \eqref{eq:HQ_master_rcless}.
The form of the master equation is different only in $D(\vec x-\vec y)t^a\hat\rho(t,\vec x,\vec y)t^a \to C_{\rm F}D(\vec x-\vec y)\rho(t,\vec x,\vec y)$.
The Lindblad operator is $\bm{L}_{\vec k}=e^{i\vec k\cdot\vec {\bm x}}$ and the coefficient is $\gamma_{\vec k}= - C_{\rm F}\tilde D(\vec k)/L^3>0$.

\subsection{Heavy quarkonium master equations}
\subsubsection{Full master equation}
In the case of quarkonium, the reduced density matrix $\hat\rho_{QQ_c}(t,\vec x_Q,\vec x_{Q_c},\vec y_Q,\vec y_{Q_c})$ is in the $(N_{\rm c}\otimes N_{\rm c}^*)\otimes(N_{\rm c}^*\otimes N_{\rm c})$ representation.
The master equation has the following structure:
\begin{eqnarray}
\label{eq:QQc_master_full}
&&\frac{\partial}{\partial t}\hat\rho_{QQ_c}(t,\vec x_Q,\vec x_{Q_c},\vec y_Q,\vec y_{Q_c})\\
&& \ \ \ =\mathcal{L}_{QQ_c}\hat\rho_{QQ_c}(t,\vec x_Q,\vec x_{Q_c},\vec y_Q,\vec y_{Q_c}), \nonumber\\
\label{eq:QQc_superop_full}
&&\mathcal{L}_{QQ_c}=\mathcal{L}_{Q}+\mathcal{L}_{Q_c}+\mathcal{L}^{(2)}_{QQ_c}.
\end{eqnarray}
Here, $\mathcal{L}_Q$ denotes the superoperator in the right-hand side of Eq.~\eqref{eq:HQ_master_full} and $\mathcal{L}_{Q_c}$ is obtained by substituting $-t^{a*}$ for $t^a$ in $\mathcal{L}_Q$.
$\mathcal{L}_{Q}$ acts on variables of heavy quark while $\mathcal{L}_{Q_c}$ acts on those of heavy antiquark.
The interaction between the heavy quark and antiquark is given by $\mathcal{L}^{(2)}_{QQ_c}$, whose explicit form is shown in Appendix \ref{sec:2body_Liouville}.

The structure of the master equation is quite complicated but the Lindblad operators and coefficients turn out to be remarkably simple.
We just need to add a contribution from a heavy antiquark with the appropriate color representation in Eqs.~\eqref{eq:HQ_Lindblad_a1}-\eqref{eq:HQ_Lindblad_a2}:
\begin{eqnarray}
&&\left\{
\begin{array}{l}
\bm{L}^{\alpha=1}_{\vec k a}=
e^{i\vec k\cdot\vec {\bm x}_Q/2}
\left(1-\frac{\vec k\cdot\vec{\bm p}_Q}{4MT}\right)
e^{i\vec k\cdot\vec {\bm x}_Q/2}(\bm t^a\otimes \bm 1)\\
\ \ \ \
-e^{i\vec k\cdot\vec {\bm x}_{Q_c}/2}
\left(1-\frac{\vec k\cdot\vec{\bm p}_{Q_c}}{4MT}\right)
e^{i\vec k\cdot\vec {\bm x}_{Q_c}/2}(\bm 1\otimes\bm t^{a*}), \\
\gamma^{\alpha=1}_{\vec k a}=-\frac{\tilde D(\vec k)}{L^3} > 0,
\end{array}\right. \ \ \ \\
&&\left\{
\begin{array}{l}
\bm{L}^{\alpha=2}_{\vec k a}=
e^{i\vec k\cdot\vec {\bm x}_Q/2}
\left(\frac{\vec k\cdot\vec{\bm p}_Q}{4MT}\right)
e^{i\vec k\cdot\vec {\bm x}_Q/2}(\bm t^a\otimes\bm 1)\\
\ \ \ \
-e^{i\vec k\cdot\vec {\bm x}_{Q_c}/2}
\left(\frac{\vec k\cdot\vec{\bm p}_{Q_c}}{4MT}\right)
e^{i\vec k\cdot\vec {\bm x}_{Q_c}/2}(\bm 1\otimes\bm t^{a*}), \\
\gamma^{\alpha=2}_{\vec k a}=-\frac{1}{L^3}
\left(8T^2\tilde A(\vec k)-\tilde D(\vec k)\right) > 0.
\end{array}\right.
\end{eqnarray}
Here, $\vec{\bm x}_{Q},\vec{\bm p}_{Q}$ are position and momentum operators for the heavy quark and $\vec{\bm x}_{Q_c},\vec{\bm p}_{Q_c}$ are those for the heavy antiquark.
The Hamiltonian in the Lindblad form \eqref{eq:Lindblad2} has two contributions in the potential:
one is the screened potential from $S_{\rm pot}$ and the other is from $S_{\rm diss}$ in the influence functional.
\begin{eqnarray}
\label{eq:QQc_Hamiltonian_full}
\bm H &=& \frac{\vec{\bm p}_Q^2+\vec{\bm p}^2_{Q_c}}{2M}
-V(\vec{\bm x}_Q-\vec{\bm x}_{Q_c})(\bm t^a\otimes\bm t^{a*})\\
&+&\frac{1}{8MT}\left\{
(\vec{\bm p}_Q-\vec{\bm p}_{Q_c}),
\vec \nabla D(\vec{\bm x}_Q - \vec{\bm x}_{Q_c})
\right\}(\bm t^a\otimes\bm t^{a*}).\nonumber
\end{eqnarray}
Note that this Hamiltonian contains a term which is time-reversal odd.
The physical meaning of the second line of Eq.~\eqref{eq:QQc_Hamiltonian_full} is remarkable.
In the classical Hamiltonian, the anticommutator part is positive (negative) when $(\vec x_Q-\vec x_{Q_c})\cdot(\vec p_Q-\vec p_{Q_c})$ is positive (negative) because $D(\vec r)$ is an increasing function of $r$.
Therefore when a heavy quark-antiquark pair in the singlet state is moving apart from (approaching) each other, the term makes a positive (negative) contribution to the Hamiltonian, while the sign is opposite for a heavy quark-antiquark pair in the octet states.

\subsubsection{Master equation in the recoilless limit}
Suppose there is a quarkonium initial state at rest in the quark-gluon plasma.
Let us parametrize the coherence length of the quarkonium bound state by $l_{\rm coh}$.
To analyze the two-body problem, it is convenient to introduce the center of mass and relative coordinates:
\begin{eqnarray}
&&\left\{
\begin{array}{l}
\vec R = \frac{\vec x_Q + \vec x_{Q_c}}{2}, \ \ \ \ \ \vec S = \frac{\vec y_Q + \vec y_{Q_c}}{2},\\
\vec r = \vec x_Q - \vec x_{Q_c}, \ \ \ \ \vec s = \vec y_Q - \vec y_{Q_c}.
\end{array}\right.
\end{eqnarray}
Since we are mainly interested in the relative motion of the heavy quark and antiquark, we take $\vec R=\vec S$.
Then by repeating the similar argument previously made, the decoherence of the wave function is the dominant process if $\Delta r\gg v_{Q,Q_c}/T$.
Here, $\Delta r$ denotes the coherence length in the relative coordinate, which is given by the typical values of $|\vec r-\vec s|$ in the wave functions (at $\vec R=\vec S$).
Typically, $|\vec r|, \ |\vec s|\alt l_{\rm coh}$ in the initial wave function and thus $\Delta r\simeq l_{\rm coh}$ holds.
Initially, the center-of-mass motion is almost static so that $v_{Q,Q_c}\simeq v_{\rm rel}/2$.
Here $v_{\rm rel}\sim 1/Ml_{\rm coh}$ is the relative velocity of the heavy quark and antiquark in the quarkonium.
Therefore, if $l_{\rm coh}\gg l_{\rm dB}\sim 1/\sqrt{MT}$ is satisfied, the dominant process for a quarkonium at rest is decoherence.
Note that $l_{\rm coh}\sim 1/M\alpha \gg l_{\rm dB}$ is satisfied by all the bound states if the condition for the coarse graining in time $M\alpha^2\ll gT$ is satisfied.

When studying the decoherence of bound states with $l_{\rm coh}\gg l_{\rm dB}$, the master equation can be approximated by keeping the kinetic term and terms from $S_{\rm pot}$ and $S_{\rm fluct}$ in the influence functional.
The superoperator $\mathcal{L}_{QQ_c}$ in the recoilless limit is
\begin{eqnarray}
\label{eq:QQc_superop_rcless1}
\mathcal{L}_Q\hat\rho_{QQ_c}
&\simeq&i\frac{\vec\nabla_{x_Q}^2-\vec\nabla_{y_Q}^2}{2M}\hat\rho_{QQ_c}\nonumber\\
&-&D(\vec x_Q-\vec y_Q)(t^a\otimes 1)\hat\rho_{QQ_c}(t^a\otimes 1)\nonumber\\
&+&C_{\rm F}D(\vec 0)\hat\rho_{QQ_c},\\
\label{eq:QQc_superop_rcless2}
\mathcal{L}_{Q_c}\hat\rho_{QQ_c}
&\simeq&i\frac{\vec\nabla_{x_{Q_c}}^2-\vec\nabla_{y_{Q_c}}^2}{2M}\hat\rho_{QQ_c}\nonumber\\
&-&D(\vec x_{Q_c}-\vec y_{Q_c})(1\otimes t^{a*})\hat\rho_{QQ_c}(1\otimes t^{a*})\nonumber\\
&+&C_{\rm F}D(\vec 0)\hat\rho_{QQ_c},
\end{eqnarray}
and 
\begin{eqnarray}
\label{eq:QQc_superop_rcless3}
&&\mathcal{L}^{(2)}_{QQ_c}\hat\rho_{QQ_c}\nonumber\\
&& \ \ \simeq \ \left(iV(\vec x_{Q}-\vec x_{Q_c})-D(\vec x_{Q}-\vec x_{Q_c})\right)
(t^a\otimes t^{a*})\hat\rho_{QQ_c}\nonumber\\
&& \ \ - \ \left(iV(\vec y_{Q}-\vec y_{Q_c})+D(\vec y_{Q}-\vec y_{Q_c})\right)
\hat\rho_{QQ_c}(t^a\otimes t^{a*})\nonumber\\
&& \ \ + \ D(\vec x_{Q}-\vec y_{Q_c})(t^a\otimes 1)\hat\rho_{QQ_c}(1\otimes t^{a*})\nonumber\\
&& \ \ + \ D(\vec y_{Q}-\vec x_{Q_c})(1\otimes t^{a*})\hat\rho_{QQ_c}(t^a\otimes 1).
\end{eqnarray}
This master equation is in the Lindblad form with
\begin{eqnarray}
\label{eq:QQc_Lindblad_rcless}
&&\left\{
\begin{array}{l}
\bm{L}_{\vec k a}=
e^{i\vec k\cdot\vec {\bm x}_Q}(\bm t^a\otimes \bm 1)
-e^{i\vec k\cdot\vec {\bm x}_{Q_c}}(\bm 1\otimes\bm t^{a*}), \\
\gamma_{\vec k a}=-\frac{\tilde D(\vec k)}{L^3} > 0,
\end{array}\right. \ \ \ 
\end{eqnarray}
and with the Hamiltonian
\begin{eqnarray}
\label{eq:QQc_Hamiltonian_rcless}
\bm H =\frac{\vec{\bm p}_Q^2+\vec{\bm p}^2_{Q_c}}{2M}
-V(\vec{\bm x}_Q-\vec{\bm x}_{Q_c})(\bm t^a\otimes\bm t^{a*}).
\end{eqnarray}

In the master equation given by the superoperators \eqref{eq:QQc_superop_rcless1}-\eqref{eq:QQc_superop_rcless3}, the relative motion and center-of-mass motion decouple.
Note that these motions decouple only after taking the recoilless limit.
Let us define the reduced density matrix for the relative motion $\hat\rho^{r}_{QQ_c}(t,\vec r,\vec s)$,
\begin{eqnarray}
&&\hat\rho^{r}_{QQ_c}(t,\vec r,\vec s) = \int d^3Rd^3S \ \delta(\vec R-\vec S)\nonumber \\
&& \ \ \ \ \times\hat\rho_{QQ_c}(t,\vec x_Q,\vec x_{Q_c},\vec y_Q,\vec y_{Q_c}),
\end{eqnarray}
and derive a master equation for it.
The result is
\begin{eqnarray}
\label{eq:QQc_master_rcless_rel}
&&\frac{\partial}{\partial t}\hat\rho^{r}_{QQ_c}(t,\vec r,\vec s)
=\left(i\frac{\vec\nabla_r^2-\vec\nabla_s^2}{M}+2C_{\rm F}D(\vec 0)\right)\hat\rho^{r}_{QQ_c}(t,\vec r,\vec s)\nonumber\\
&& \ \ + \ (iV(\vec r)-D(\vec r))(t^a\otimes t^{a*})\hat\rho^{r}_{QQ_c}(t,\vec r,\vec s)\nonumber\\
&& \ \ - \ (iV(\vec s)+D(\vec s))\hat\rho^{r}_{QQ_c}(t,\vec r,\vec s)(t^a\otimes t^{a*})\nonumber\\
&& \ \ - \ D\left(\frac{\vec r-\vec s}{2}\right)\left(
\begin{array}{l}
(t^a\otimes 1)\hat\rho^{r}_{QQ_c}(t,\vec r,\vec s)(t^a\otimes 1)\\
+(1\otimes t^{a*})\hat\rho^{r}_{QQ_c}(t,\vec r,\vec s)(1\otimes t^{a*})
\end{array}
\right)\nonumber\\
&& \ \ + \ D\left(\frac{\vec r+\vec s}{2}\right)\left(
\begin{array}{l}
(t^a\otimes 1)\hat\rho^{r}_{QQ_c}(t,\vec r,\vec s)(1\otimes t^{a*})\\
+(1\otimes t^{a*})\hat\rho^{r}_{QQ_c}(t,\vec r,\vec s)(t^a\otimes 1)
\end{array}
\right).
\end{eqnarray}
The Lindblad operator is obtained by substituting $\vec{\bm x}_Q\to\vec{\bm r}/2$ and $\vec{\bm x}_{Q_c}\to -\vec{\bm r}/2$ in Eq.~\eqref{eq:QQc_Lindblad_rcless} and the coefficient is the same as Eq.~\eqref{eq:QQc_Lindblad_rcless}.
The Hamiltonian is obtained by just expressing Eq.~\eqref{eq:QQc_Hamiltonian_rcless} in the relative coordinate.
In Table \ref{tab:conditions_QQcME}, we summarize the recoilless limit master equation for the quarkonium.

\begin{table}[top]
\begin{tabular}{c|c}
\hline
  & Recoilless limit \\
\hline
Bound state size & $l_{\rm coh} \gg l_{\rm dB}$ \\
\hline
Approximation of $S_{\rm IF}$ &
$S_{\rm IF}\simeq S_{\rm pot} + S_{\rm fluct}$ \\
\hline
Physical process & Decoherence \\
\hline
\end{tabular}
\caption{
Summary of the approximated master equation \eqref{eq:QQc_master_rcless_rel} for the quarkonium.
}
\label{tab:conditions_QQcME}
\end{table}

Since the potential and thermal fluctuation depends on the color states of quarkonium, one cannot trace out the color space dynamics in the master equation \eqref{eq:QQc_master_rcless_rel}.
Instead, we can obtain coupled master equations for the color singlet occupation $\rho_1(t,\vec r,\vec s)\equiv {\rm Tr}_{\rm color}\left[\hat\rho^r_{QQ_c}(t,\vec r,\vec s)P_{1}\right]$ and for the color octet [or $(N_{\rm c}^2-1)$ representation] occupation $\rho_8(t,\vec r,\vec s)\equiv {\rm Tr}_{\rm color}\left[\hat\rho^r_{QQ_c}(t,\vec r,\vec s)P_{8}\right]$, where $P_1$ and $P_8$ are projection operators onto color singlet and octet states.

\section{Quantum decoherence of heavy quarkonium}
\label{sec:decoherence}
In this section, we show that the master equations in the recoilless limit are equivalent to stochastic Schr\"odinger equations.
The stochastic Schr\"odinger equations describe the effects of thermal fluctuation on the quantum states of heavy quarks.
Because of the thermal fluctuation, the wave function at distant points becomes decoherent.
Decoherence is essential for quarkonium dissociation so that here we concentrate on the quarkonium sector.
Since the stochastic Schr\"odinger equations can be understood as Hamiltonian dynamics with time-dependent random potential, they cannot describe irreversible processes such as momentum dissipation.
We also discuss decoherence and classicalization of a wave function.
After the system enters in the classical regime, classical descriptions, such as \cite{Young:2008he}, would become applicable.

\subsection{Stochastic potential}
The basics of the stochastic potential are given in \cite{Akamatsu:2011se}.
The wave function of a quarkonium $\psi(t,\vec x_Q,\vec x_{Q_c})$ is in the $N_{\rm c}\otimes N_{\rm c}^*$ representation.
The stochastic and unitary time evolution of $\psi(t,\vec x_Q,\vec x_{Q_c})$ is
\begin{eqnarray}
\psi(t+dt,\vec x_Q,\vec x_{Q_c})=e^{-idt H_{\theta}(t)}
\psi(t,\vec x_Q,\vec x_{Q_c}),
\end{eqnarray}
with the following stochastic Hamiltonian:
\begin{eqnarray}
&&H_{\theta}(t)=H + \theta^a(t,\vec x_Q)(t^a\otimes 1) - \theta^a(t,\vec x_{Q_c})(1\otimes t^{a*}),\nonumber \\
&&H=-\frac{\vec{\nabla}_{x_Q}^2+\vec{\nabla}_{x_{Q_c}}^2}{2M}
-V(\vec x_Q - \vec x_{Q_c})(t^a\otimes t^{a*}), \nonumber\\
&&\langle \theta^a(t,\vec x)\theta^b(s,\vec y)\rangle = -D(\vec x-\vec y)\delta(t-s)\delta^{ab}.
\end{eqnarray}
Note that $-D(\vec r)$ is positive definite.
In the limit $dt\to 0$, the stochastic Schr\"odinger equation becomes (in the It$\hat {\rm o}$ discretization)
\begin{eqnarray}
\label{eq:QQc_stochastic}
&&i\frac{\partial}{\partial t}\psi(t,\vec x_Q,\vec x_{Q_c})
=H_{\xi}(t)\psi(t,\vec x_Q,\vec x_{Q_c}),\\
&&H_{\xi}(t)=H_{\theta}(t)+iC_{\rm F}D(\vec 0)-iD(\vec x_Q-\vec x_{Q_c})(t^a\otimes t^{a*}).\nonumber\\
\end{eqnarray}
In the stochastic Hamiltonian $H_{\xi}(t)$, we omit terms of the form $dt (\theta^2-\langle\theta^2\rangle)\sim \mathcal O(dt^0)$ because they do not contribute in the master equation.
In the stochastic Schr\"odinger equation, the reduced density matrix is defined as $\hat\rho_{QQ_c}(t,\vec x_Q,\vec x_{Q_c},\vec y_Q,\vec y_{Q_c}) \equiv \langle \psi(t,\vec x_Q,\vec x_{Q_c})\psi^*(t,\vec y_Q,\vec y_{Q_c})\rangle_{\theta}$ and its time evolution is governed by the master equation obtained previously.

In the relative coordinate, the stochastic Schr\"odinger equation for the wave function $\psi^r (t,\vec r)$ is also obtained similarly:
\begin{eqnarray}
\label{eq:QQc_stochastic_rel}
&&i\frac{\partial}{\partial t}\psi^r(t,\vec r)
=H^r_{\xi}(t)\psi^r(t,\vec r),\\
\label{eq:QQc_stochastic_rel_Hamiltonian}
&&H^r_{\xi}(t)= -\frac{\vec{\nabla}_r^2}{M}+iC_{\rm F}D(\vec 0)
+\left(-V(\vec r)-iD(\vec r)\right)(t^a\otimes t^{a*})\nonumber\\
&&\ \ \ \ \ \ \ \ + \ \theta^a(t,\vec r/2)(t^a\otimes 1) - \theta^a(t,-\vec r/2)(1\otimes t^{a*}),
\end{eqnarray}
and the master equation \eqref{eq:QQc_master_rcless_rel} is obtained by defining the reduced density matrix $\hat\rho^r_{QQ_c}(t,\vec r,\vec s)\equiv\langle \psi^r(t,\vec r)\psi^{r*}(t,\vec s)\rangle_{\theta}$.

In a numerical simulation, solving the stochastic Schr\"odinger equation has a substantial advantage over solving the master equation because the dimension of the former is the square root of the latter.

\subsection{Heavy quarkonium dissociation}
In the stochastic Schr\"odinger equations \eqref{eq:QQc_stochastic} and \eqref{eq:QQc_stochastic_rel}, the noise and imaginary part describes how the color density fluctuation in the medium affects quantum dynamics while the potential describes how the heavy quark and antiquark interact with each other in the medium.
The important scales here are correlation length $l_{\rm fluct}$ of the color density fluctuation $-D(\vec r)$ and the range of the screened potential $V(\vec r)$ (or more precisely coherence length $l_{\rm coh}$ of the bound states).
If the former is much longer than the latter $l_{\rm fluct}\gg l_{\rm coh}\gg l_{\rm dB}$,
the wave function remains almost unchanged by a scattering except for receiving a nearly uniform but random phase factor.
In the opposite case $l_{\rm fluct} \alt l_{\rm coh}$, the wave function easily becomes decoherent by a scattering.

To see these features explicitly, let us write down the coupled master equations for density matrices projected onto color singlet and octet states [$\rho_1(t,\vec r,\vec s)$ and $\rho_8(t,\vec r,\vec s)$ defined previously]:
\begin{eqnarray}
&&\frac{\partial}{\partial t}
\left(
\begin{array}{c}
\rho_1 \\
\rho_8
\end{array}
\right)_{(t,\vec r,\vec s)}
=\left(
i\frac{\vec{\nabla}_r^2-\vec{\nabla}_s^2}{M}
\right)
\left(
\begin{array}{c}
\rho_1 \\
\rho_8
\end{array}
\right)_{(t,\vec r,\vec s)} \nonumber \\
&& \ \ \ \ \ + \ i\left(
V(\vec r)-V(\vec s)
\right)
\left[
\begin{array}{cc}
C_{\rm F} & 0 \\
0  & -1/2N_{\rm c}
\end{array}
\right]
\left(
\begin{array}{c}
\rho_1 \\
\rho_8
\end{array}
\right)_{(t,\vec r,\vec s)}\nonumber \\
&& \ \ \ \ \ + \ \mathcal D(\vec r,\vec s)
\left(
\begin{array}{c}
\rho_1 \\
\rho_8
\end{array}
\right)_{(t,\vec r,\vec s)},
\end{eqnarray}
where $\mathcal D(\vec r,\vec s)$, which describes decoherence, is defined as
\begin{eqnarray}
\mathcal D(\vec r,\vec s)
&=&2C_{\rm F}D(\vec 0)
-\left(
D(\vec r)+D(\vec s)
\right)
\left[
\begin{array}{cc}
C_{\rm F} & 0 \\
0  & -1/2N_{\rm c}
\end{array}
\right]\nonumber\\
&-&2D\left(
\frac{\vec r-\vec s}{2}
\right)
\left[
\begin{array}{cc}
0 & 1/2N_{\rm c} \\
C_{\rm F}  & C_{\rm F}-1/2N_{\rm c}
\end{array}
\right]\nonumber \\
&+&2D\left(
\frac{\vec r+\vec s}{2}
\right)
\left[
\begin{array}{cc}
0 & 1/2N_{\rm c} \\
C_{\rm F}  & -1/N_{\rm c}
\end{array}
\right].
\end{eqnarray}
Before discussing decoherence, let us start from a simpler case with $\vec r = \vec s$ as a warm-up.
Since $\rho_1(t,\vec r,\vec r)$ and $\rho_8(t,\vec r,\vec r)$ represent probability densities to find a quarkonium with separation $\vec r$ in the singlet and octet states, $\mathcal D(\vec r,\vec s)$ at the same points $\vec r=\vec s$ gives the rate of color singlet-octet transitions there:
\begin{eqnarray}
\mathcal D(\vec r,\vec r)=2(D(\vec 0)-D(\vec r))
\left[
\begin{array}{cc}
C_{\rm F} & -1/2N_{\rm c} \\
-C_{\rm F}  & 1/2N_{\rm c}
\end{array}
\right].
\end{eqnarray}
Because $D(\vec 0)-D(\vec r) < 0 \ (|\vec r| \neq 0)$, $\mathcal D(\vec r,\vec r)$ has zero and negative eigenvalues for eigenvectors $^t(1, N_{\rm c}^2-1)$ and $^t(1,-1)$.
The eigenvector $^t(1, N_{\rm c}^2-1)$ represents the equal occupation in the color singlet and octet states.
If we ignore the kinetic and potential terms, a color space configuration would approach this state, within a shorter time scale at larger $|\vec r|$.
It is also important to observe that $\rho_1(t,\vec r,\vec r)+\rho_8(t,\vec r,\vec r)$ is conserved by $\mathcal D(\vec r,\vec r)$.
In the recoilless limit, the scatterings are equivalently described by a stochastic potential, which randomly gives phase and color rotations to a wave function.
Therefore, the probability density to find a quarkonium with a given separation $\vec r$ either in the color singlet or octet states must be conserved in each scattering in the recoilless limit.

Now let us discuss the decoherence of a wave function.
If the coherence length of a wave function is small $l_{\rm fluct}\gg l_{\rm coh}\gg l_{\rm dB}$, we have $D(\vec r),D(\vec s)\simeq D(\vec 0)$ for $|\vec r|,|\vec s| \simeq l_{\rm coh}$ in the domain of the wave function.
In this case, the decoherence is not effective $\mathcal D(\vec r,\vec s)\simeq 0$.
Note that this holds both for $\rho_1(t,\vec r,\vec s)$ and $\rho_8(t,\vec r,\vec s)$ even though the color singlet and octet states are quite different in their interaction with medium particles:
Since the wave function is localized, the singlet state is almost invisible to them while the octet states clearly interact with them.
An octet state does interact with the medium but it remains as one of the octet states.
In the limit of small wave function $|\vec r|,|\vec s|\to 0$, the octet state can be regarded as a pointlike gluon so that it remains octet through the interaction.
\footnote{
More specifically, one can see that an operator $\theta^a(t,\vec 0)\left[(t^a\otimes 1)-(1\otimes t^{a*})\right]$ in Eq.~\eqref{eq:QQc_stochastic_rel_Hamiltonian} maps an octet state to another octet state.
The singlet state is a zero mode of this operator.
}
This is why $\mathcal D(\vec r,\vec s)\simeq 0$ also for the octet states.
This kind of information cannot be gained just from the imaginary part of the potential.
By taking $\vec s\simeq -\vec r$ and expanding $\mathcal D(\vec r, -\vec r)$ in terms of $|\vec r|/l_{\rm fluct}\simeq l_{\rm coh}/l_{\rm fluct}\ll 1$ to second order, we obtain
\begin{eqnarray}
\mathcal D(\vec r,-\vec r)\simeq -\frac{D_2\vec r^2}{3}
\left[
\begin{array}{cc}
C_{\rm F} & 1/2N_{\rm c} \\
C_{\rm F}  & C_{\rm F}-1/N_{\rm c}
\end{array}
\right].
\end{eqnarray}
Here, $\mathcal D(\vec r,-\vec r)$ has only negative eigenvalues and the time scale of decoherence at the opposite edges of the wave function is estimated as $\sim 1/D_2l_{\rm coh}^2\sim [g^4\ln(1/g)T^3l_{\rm coh}^2]^{-1}$. 

If the coherence length of a wave function is large $l_{\rm coh}\agt l_{\rm fluct}$, we have $D(\vec r),D(\vec s)\simeq 0$ for $|\vec r|,|\vec s| \simeq l_{\rm coh}\agt l_{\rm fluct}$ and thus the decoherence at the edges of the wave function $\vec s\simeq -\vec r$ is given by
\begin{eqnarray}
\mathcal D(\vec r,-\vec r)\simeq 2D(\vec 0)
\left[
\begin{array}{cc}
C_{\rm F} & 1/2N_{\rm c} \\
C_{\rm F}  & C_{\rm F}-1/N_{\rm c}
\end{array}
\right].
\end{eqnarray}
Because $D(\vec 0)<0$, $\mathcal D(\vec r,-\vec r)$ has only negative eigenvalues so that it makes the wave function decoherent by scatterings.
The time scale for the decoherence is $\sim 1/D(\vec 0)\sim 1/g^2T$.
In this regime, the potential $V(\vec r)$ is screened and does not play an important role.

In summary, we have shown that the bound states with larger size dissociate more easily by scattering with medium particles, as one can imagine quite intuitively.
We can simply parametrize the decoherence time scale by
\begin{eqnarray}
\label{eq:decoh_timescale}
t_{\rm D}(l_{\rm coh},T)\sim \frac{1}{g^2T}\left(a+\frac{b}{g^2\ln (1/g)T^2l_{\rm coh}^2}\right),
\end{eqnarray}
with dimensionless coefficients $a$ and $b$ of order $\mathcal O(g^0)$.
Comparison of the decoherence time scale $t_{\rm D}$ and the lifetime of the QGP fireball in heavy-ion collisions will give us a rough estimate of quarkonium dissociation.
For detailed information, such as the occupation number of a state at a given time, we need to solve the master equation or its equivalent stochastic Schr\"odinger equation.
In particular, there is the non-negligible probability that the octet states get deexcited to the singlet bound states in the medium.
This process is not captured by the decoherence time scale.

\subsection{Classicalization}
\label{sec:classicalization}
When the quarkonium wave function becomes decoherent and typically $|\vec r|,|\vec s|\agt l_{\rm fluct}$, the medium interacts with the heavy quark and antiquark independently.
However, even after the wave function becomes decoherent, the wave property still remains until the medium correlation length $l_{\rm fluct}$ cannot resolve the wave packet of size $\Delta r$.
Note that the wave packet here has color in the fundamental representation.

Once the wave packet becomes small enough compared to $l_{\rm fluct}\gg \Delta r$, a classical description is applicable and practically suitable.
\footnote{
We may also call this process decoherence.
The difference is whether it is for one-body or two-body wave functions.
}
For example, one can obtain phase space distribution of the heavy quark and antiquark by Wigner transformation and switch to classical description such as Ref.~\cite{Young:2008he}.
Since the master equation in the recoilless limit is applicable for $\Delta r\gg l_{\rm dB}$, there exists a regime $l_{\rm fluct}\gg \Delta r\gg l_{\rm dB}$ where the switch to classical description is possible.

\section{Summary}
\label{sec:summary}
In this paper, we have derived the Lindblad-form master equations for heavy quark systems in the quark-gluon plasma.
The master equation in the Lindblad form ensures the complete positivity of the reduced density matrix as it evolves in time.
Therefore, deriving the master equations in the Lindblad form is an important theoretical advance in the formulation of quantum dynamics of heavy quarks.

In order to obtain the master equations in the Lindblad form, we derive the influence functional $S_{\rm IF}$ by perturbative expansion and by coarse graining in time.
In the heavy quark Lagrangian, we take the nonrelativistic limit and keep the leading terms in the $1/c$ expansion, namely, the color density interaction terms.
The influence functional consists of $S_{\rm pot},S_{\rm fluct}\sim \mathcal O(g^2v^0)$,  $S_{\rm diss}\sim \mathcal O(g^2v)$, and $S_{\rm L}\sim\mathcal O(g^2v^2)$.
Here $S_{\rm L}$ plays an essential role in deriving the master equations in the Lindblad form.
The velocity $v$ comes into play in the course of the coarse graining in time.

In the coarse graining, we need a condition $M\alpha^2 \ll gT$ in order to neglect the effect of acceleration in the quarkonium bound states during a scattering event.
This regime is called the quantum Brownian motion in the open quantum systems.
When $M\alpha^2 \ll gT$ is not satisfied, it indicates that quantum optical description works better for a quarkonium.
In such a case, the master equations for a single heavy quark and those for a quarkonium are not derived from a common influence functional $S_{\rm IF}$.

After deriving the master equations in the Lindblad form, we have made approximations to obtain more effective master equations appropriate to the physical conditions of the problems.
One is for Langevin dynamics of localized wave packets and the other is for the decoherece of extended wave functions.
Both approximations yield master equations in the Lindblad form.

Finally, we have examined the decoherence of a quarkonium wave function.
The decoherence is described by the master equation in the recoilless limit, which is equivalent to the Schr\"odinger equation with a stochastic potential.
In terms of a stochastic potential, quarkonium dissociation can be understood as an interplay of two length scales, the coherence length of a state $l_{\rm coh}$ and the correlation length of the thermal fluctuation $l_{\rm fluct}$.
For $ l_{\rm fluct} \gg l_{\rm coh}\gg l_{\rm dB}$, the decoherence of the wave function is not effective and quarkonium dissociation requires a longer time of the order $\sim [g^4\ln(1/g)T^3l_{\rm coh}^2]^{-1}$.
For $l_{\rm fluct} \alt l_{\rm coh}$, the decoherence is so efficient that the quarkonium dissociates quickly with the typical time scale $\sim 1/g^2T$.
The recoilless limit master equation can also describe the classicalization until the wave packet size $\Delta r$ becomes too small to be resolved by the medium fluctuation $l_{\rm fluct}\gg \Delta r\gg l_{\rm dB}$.
In this regime, one can switch to a classical description that is more effective.

As future prospects, the calculation of the $\Upsilon$ spectrum at the LHC is one of the important applications of our approach.
For this application, we need to model the dynamics of the open quantum system in the nonperturbative region by referring to and extending the perturbative results.
It is also an open problem to describe the real gluon processes, such as excitation of quarkonium by absorbing a real gluon (gluodissociation).
For this extension, we need to keep higher order terms in the $1/c$ expansion or the $1/M$ expansion in the heavy quark Lagrangian.

\section*{ACKNOWLEDGEMENTS}
I thank Fran\c{c}ois Gelis and Derek Teaney for enlightening discussions during their stays at the Kobayashi-Maskawa Institute as KMI visitors.
I also thank Jean-Paul Blaizot, Tetsuo Hatsuda, and Alexander Rothkopf for valuable comments on the manuscript.
Finally, I thank the Institute for Nuclear Theory at the University of Washington for their kind hospitality at ``Heavy Flavor and Electromagnetic Probes in Heavy Ion Collisions,'' during which part of the revision was made.

\appendix
\section{TWO-POINT FUNCTIONS OF GLUONS}
\label{sec:2point_functions}
The influence functional $S_{\rm IF}$ up to the order of $\mathcal O(g^2v^0,g^2v)$ and some of $\mathcal O(g^2v^2)$ is given by two-point functions of gluons.
Since we are interested in the distance scale of $r\simeq 1/gT$, where the Debye screening of the color charges becomes important, we need to include HTL resummations to obtain the two-point functions at the leading order $\mathcal O(g^2)$.

The two-point functions $V(\vec r)$, $D(\vec r)$, and $A(\vec r)$ are defined by using the retarded propagator $\tilde G^{\rm R}_{ab,00}(\omega,\vec r)$ and the spectral function $\sigma_{ab,00}(\omega,\vec r)$, as shown in Eqs.~\eqref{eq:potential}-\eqref{eq:new}.
The explicit forms of the retarded propagator and the spectral function are
\begin{eqnarray}
\label{eq:GR}
&&\tilde G^{\rm R}_{ab,00}(0,\vec r)
=-\frac{e^{-\omega_{\rm D}r}}{4\pi r},\\
\label{eq:sigma1}
&&\frac{\partial}{\partial \omega}\sigma_{ab,00}(0,\vec r)
=\int\frac{d^3k}{(2\pi)^3}
\frac{\pi\omega_{\rm D}^2e^{i\vec k\cdot\vec r}}{k(k^2+\omega_{\rm D}^2)^2}, % \\
%\label{eq:sigma3}
%&&\frac{\partial^3}{\partial \omega^3}\sigma_{ab,00}(0,\vec r)
%=\int\frac{d^3k}{(2\pi)^3}
%\frac{3\pi\omega_{\rm D}^2e^{i\vec k\cdot\vec r}}{k(k^2+\omega_{\rm D}^2)^2}\\
%&& \ \ \ \ \ \ \ \ \ \ \
%\times
%\left(
%\frac{4}{k^2}+\frac{4\omega_{\rm D}^2}{k^2(k^2+\omega_{\rm D}^2)}
%-\frac{\pi^2\omega_{\rm D}^4}{2k^2(k^2+\omega_{\rm D}^2)^2}
%\right),\nonumber
\end{eqnarray}
with the Debye screening mass $\omega_{\rm D}^2=(g^2T^2/3)(N_{\rm c}+N_{\rm f}/2)$ for QCD with $N_{\rm f}$ light flavors.
With these, $V(\vec r)$, $D(\vec r)$, and $A(\vec r)$ are determined to leading order in $g$.

It should be emphasized that the HTL-resummed calculation gives the leading-order result for $r\simeq 1/gT$ but does not give a correct extrapolation from $r\simeq 1/gT$ to $r\simeq 1/T$.
For example, if we calculate the heavy quark momentum diffusion constant, which is given by $(C_{\rm F}/3)\vec \nabla^2 D(\vec r)|_{r=0}$ \cite{Akamatsu:2012vt}, the scattering processes with exchanged momentum $k\simeq T$ (hard) as well as $k\simeq gT$ (soft) become relevant.
In this case, we need to split the momentum integral at some intermediate scale $gT\ll\Lambda\ll T$ in momentum space and add the two contributions to obtain the heavy quark diffusion constant.
In $k<\Lambda$ the HTL-resummed result for soft momentum exchange is reliable, while in $k>\Lambda$ the scattering processes with hard exchanged momentum $k$ need to be considered separately.
The contributions from different momentum regions are logarithmically sensitive to the scale $\Lambda$ but these dependences are canceled in the sum, yielding a finite and $\Lambda$-independent heavy quark momentum diffusion constant \cite{Moore:2004tg}.

\section{AN ILLUSTRATION OF OBTAINING THE MASTER EQUATIONS}
\label{sec:illustration}
Here we briefly sketch how a term
$-\frac{1}{2}\int_{t^0}dt\int d^3xd^3y D(\vec x-\vec y)\rho^a_1(t,\vec x)\rho^a_2(t,\vec y)\in iS_{\rm fluct}$
in the influence functional contributes in the master equation for a single heavy quark.
First, we obtain the corresponding term in the Hamiltonian ${\bm H}_{\rm CTP}$ as
\begin{eqnarray}
{\bm H}_{\rm CTP}&\ni& \frac{i}{2}\int_{t^0}dt\int d^3xd^3y D(\vec x-\vec y)[t^a]_{ij}[t^a]_{kl}\\
&& \ \ \ \ \ \ \ \times \ {\bm\psi}_{1}^{i\dagger}(t,\vec x) {\bm \psi}_{1}^j(t,\vec x)
{\bm \psi}_{2}^l(t,\vec y){\bm\psi}_{2}^{k\dagger}(t,\vec y).\nonumber
\end{eqnarray}
Note that the operators are ordered by time.
The overall sign is determined by $(-1)^2$: $(-1)$ from conversion to Hamiltonian and $(-1)$ from fermion field ordering for $\psi_2$.
In the single heavy quark sector, the relevant term is
\begin{eqnarray}
{\bm H}_{\rm CTP}&\ni& \frac{i}{2}\int_{t^0}dt\int d^3xd^3y D(\vec x-\vec y)[t^a]_{ij}[t^a]_{kl} \\
&& \ \ \ \ \ \ \ \times \ {\bm Q}_{1}^{i\dagger}(t,\vec x) {\bm Q}_{1}^j(t,\vec x)
{\bm Q}_{2}^l(t,\vec y){\bm Q}_{2}^{k\dagger}(t,\vec y).\nonumber
\end{eqnarray}
We introduce $(\tilde {\bm Q}_2^{\dagger},\tilde {\bm Q}_2)\equiv ({\bm Q}_2,{\bm Q}_2^{\dagger})$ and obtain a functional operator by replacing
$({\bm Q}_1^{\dagger}, {\bm Q}_1)\to (Q_1^*, \frac{\delta}{\delta Q^*_1})$ and 
$(\tilde{\bm Q}_2^{\dagger}, \tilde {\bm Q}_2)\to (\tilde Q_2^*, -\frac{\delta}{\delta \tilde Q^*_2})$.
This yields
\begin{eqnarray}
H_{\rm CTP}&\ni& -\frac{i}{2}\int_{t^0}dt\int d^3xd^3y D(\vec x-\vec y)[t^a]_{ij}[t^a]_{kl} \\
&& \ \ \ \ \ \ \ \times \ Q_{1}^{i*}(t,\vec x) \frac{\delta}{\delta Q_{1}^{j*}(t,\vec x)}
\tilde Q_{2}^{l*}(t,\vec y)\frac{\delta}{\delta \tilde Q_{2}^{k*}(t,\vec y)}\nonumber \\
&=& -\frac{i}{2}\int_{t^0}dt\int d^3xd^3y D(\vec x-\vec y)[t^a]_{ij}[t^{a*}]_{lk} \nonumber \\
&& \ \ \ \ \ \ \ \times \ \tilde Q_{2}^{l*}(t,\vec y) Q_{1}^{i*}(t,\vec x) 
\frac{\delta}{\delta Q_{1}^{j*}(t,\vec x)}\frac{\delta}{\delta \tilde Q_{2}^{k*}(t,\vec y)}.\nonumber
\end{eqnarray}
This functional operator acts on $\rho_{\rm S}[t,Q_{1(c)}^*,\tilde Q_{2(c)}^*]$ in the functional master equation $i\partial_t\rho_{\rm S} = H_{\rm CTP}\rho_{\rm S}$.
As in Eq.\eqref{eq:particle_density_matrix}, reduced density matrix of a single heavy quark is given by
\begin{eqnarray}
\rho^{ij}_{Q}(t,\vec x,\vec y)
&=&-\frac{\delta}{\delta Q^{i*}_1(\vec x)}\frac{\delta}{\delta \tilde Q_2^{j*}(\vec y)}
\rho_{\rm S}\left[t,Q^*_{1(c)},\tilde Q_{2(c)}^*\right]\Big |_{Q^*=0}.\nonumber \\
\end{eqnarray}
Thus, we obtain the master equation for a single heavy quark as
\begin{eqnarray}
&&\frac{\partial}{\partial t}\rho^{ij}_Q(t,\vec x,\vec y)\\
&&=\left[\cdots - \frac{1}{2}D(\vec x-\vec y)[t^a]_{ik}[t^{a*}]_{jl}+\cdots\right]\rho^{kl}_Q(t,\vec x,\vec y).\nonumber
\end{eqnarray}
We have the same contribution from $-\frac{1}{2}\int_{t^0}dt\int d^3xd^3y D(\vec x-\vec y)\rho^a_2(t,\vec x)\rho^a_1(t,\vec y)\in iS_{\rm fluct}$ so that we see the sum of these in the master equation \eqref{eq:HQ_master_full} (the first term in $F_1$).

\section{TIME-EVOLUTION EQUATION FOR THE FORWARD PROPAGATOR}
\label{sec:forward_propagator}
By the method explained in Sec.~\ref{sec:influence_functional}, we can derive the time-evolution equation for the forward propagator $G^{>}_{QQ_c}(t,\vec x_Q,\vec x_{Q_c})$ in the $N_{\rm c}\otimes N_{\rm c}^*$ representation.
The time-evolution equation is often called the Schr\"odinger equation, causing a lot of confusion by its name.
Using the influence functional up to $\mathcal O(g^2v^0,g^2v)$ [thus we do not consider $S_{\rm L}\sim\mathcal O(g^2v^2)$ here], the time evolution of the forward propagator is given by an operator $K(\vec x_Q,\vec x_{Q_c})$:
\begin{eqnarray}
\label{eq:QQc_forward}
&&i\frac{\partial}{\partial t}G^{>}_{QQ_c}(t,\vec x_Q,\vec x_{Q_c})
=K(\vec x_Q,\vec x_{Q_c})G^{>}_{QQ_c}(t,\vec x_Q,\vec x_{Q_c}),\nonumber\\
\ \\
\label{eq:QQc_evolution}
&&K(\vec x_Q,\vec x_{Q_c})=\left\{
\begin{array}{l}
2M-\frac{\vec{\nabla}_{x_Q}^2+\vec{\nabla}^2_{x_{Q_c}}}{2M}\\
+C_{\rm F}\left(-\frac{g^2\omega_D}{4\pi}+iD(\vec 0)+i\frac{\vec\nabla^2D(\vec 0)}{4MT} \right)
\end{array}
\right\}\nonumber\\
&& \ \ \
+\left\{
\begin{array}{l}
-V(\vec{x}_Q-\vec{x}_{Q_c})
-iD(\vec{x}_Q-\vec{x}_{Q_c})\\
-\frac{i}{4MT}
\left(
\begin{array}{l}
\vec{\nabla}_{x_Q}^2 D(\vec{x}_Q-\vec{x}_{Q_c})\\
+\vec\nabla_{x_Q} D(\vec{x}_Q-\vec{x}_{Q_c})\cdot (\vec \nabla_{x_Q}-\vec \nabla_{x_{Q_c}})
\end{array}
\right)
\end{array}
\right\}\nonumber\\
&& \ \ \ \ \ \ \ 
\times(t^a\otimes t^{a*}).
\end{eqnarray}
We find that there are terms not only from $S_{\rm pot}, S_{\rm fluct}\sim\mathcal O(g^2v^0)$ but also from $S_{\rm diss}\sim\mathcal O(g^2v)$ in the operator $K(\vec x_Q,\vec x_{Q_c})$.
This shows that the leading correction to the operator $K(\vec x_Q,\vec x_{Q_c})$ in the velocity expansion is $\mathcal O(v)$.
The $\mathcal O(v)$ term comes from the diagonal parts of $S_{\rm diss}$.
This is correctly obtained by choosing $t={\rm max}(x^0,y^0)$.

In Eq.~\eqref{eq:QQc_evolution}, the term $-\frac{i}{4MT}(\cdots)(t^a\otimes t^{a*})$ in $K(\vec x_Q,\vec x_{Q_c})$ is Hermitian and identical to the second line of Eq.~\eqref{eq:QQc_Hamiltonian_full}.
Therefore, by projecting Eqs.~\eqref{eq:QQc_forward}-\eqref{eq:QQc_evolution} onto the singlet channel, we can see that the term makes a positive (negative) contribution to $K(\vec x_Q,\vec x_{Q_c})$ when a heavy quark-antiquark pair is moving apart from (approaching) each other.
The sign is opposite if we project onto the octet channel.

\section{EXPLICIT FORM OF $\mathcal{L}_{QQ_c}^{(2)}$}
\label{sec:2body_Liouville}
The explicit form of $\mathcal{L}_{QQ_c}^{(2)}$ consists of four terms:
\begin{eqnarray}
\mathcal{L}_{QQ_c}^{(2)}\hat{\rho}_{QQ_c}&=&
\mathcal{L}^{11}_{QQ_c}(\vec x_Q,\vec x_{Q_c})(t^a\otimes t^{a*})\hat{\rho}_{QQ_c}\nonumber\\
&+&\mathcal{L}^{12}_{QQ_c}(\vec x_Q,\vec y_{Q_c})(t^a\otimes 1)\hat{\rho}_{QQ_c}(1\otimes t^{a*})\nonumber\\
&+&\mathcal{L}^{21}_{QQ_c}(\vec y_Q,\vec x_{Q_c})(1\otimes t^{a*})\hat{\rho}_{QQ_c}(t^a\otimes 1)\nonumber\\
&+&\mathcal{L}^{22}_{QQ_c}(\vec y_Q,\vec y_{Q_c})\hat{\rho}_{QQ_c}(t^a\otimes t^{a*}).
\end{eqnarray}
With $\vec r_{11}=\vec x_Q-\vec x_{Q_c}$ and $\vec r_{12}=\vec x_Q-\vec y_{Q_c}$, each of them is given by
%%%%%
\begin{eqnarray}
&&\mathcal{L}^{11}_{QQ_c}(\vec x_Q,\vec x_{Q_c})\nonumber \\
&& \ \ \ =iV(\vec r_{11})-D(\vec r_{11})
-\frac{\nabla^2 D(\vec r_{11})}{4MT}
+\frac{(\vec{\nabla}^2)^2 A(\vec r_{11})}{8M^2}\nonumber\\
&& \ \ \ - \ \vec\nabla\left(
\frac{D(\vec r_{11})}{4MT}
-\frac{\vec{\nabla}^2A(\vec r_{11})}{4M^2}
\right)\cdot (\vec \nabla_{x_Q}-\vec \nabla_{x_{Q_c}})\nonumber\\
&& \ \ \ - \ \frac{\nabla^i\nabla^j A(\vec r_{11})}{2M^2}\nabla_{x_Q}^i\nabla_{x_{Q_c}}^j,
\end{eqnarray}
%%%%%
\begin{eqnarray}
&&\mathcal{L}^{12}_{QQ_c}(\vec x_Q,\vec y_{Q_c})\nonumber \\
&& \ \ \ =D(\vec r_{12})
+\frac{\nabla^2 D(\vec r_{12})}{4MT}
+\frac{(\vec{\nabla}^2)^2 A(\vec r_{12})}{8M^2}\nonumber\\
&& \ \ \ + \ \vec\nabla\left(
\frac{D(\vec r_{12})}{4MT}
+\frac{\vec{\nabla}^2A(\vec r_{12})}{4M^2}
\right)\cdot (\vec \nabla_{x_Q}-\vec \nabla_{y_{Q_c}})\nonumber\\
&& \ \ \ - \ \frac{\nabla^i\nabla^j A(\vec r_{12})}{2M^2}\nabla_{x_Q}^i\nabla_{y_{Q_c}}^j,
\end{eqnarray}
%%%%%
\begin{eqnarray}
&&\mathcal{L}^{21}_{QQ_c}(\vec y_Q,\vec x_{Q_c})
=\mathcal{L}^{12}_{QQ_c}(\vec y_Q,\vec x_{Q_c}), \\
&&\mathcal{L}^{22}_{QQ_c}(\vec y_Q,\vec y_{Q_c})
=\left(\mathcal{L}^{11}_{QQ_c}(\vec y_Q,\vec y_{Q_c})\right)^*.
\end{eqnarray}

\bibliographystyle{apsrev}

\end{document}